\documentclass[showpacs,showkeys,nofootinbib,amsfonts,amssymb,floatfix,
twocolumn,
tightenlines,superscriptaddress]{revtex4-2}
\pdfoutput=1

\usepackage{graphicx}  
\usepackage{amsmath, amssymb, bm,bbm}   % AMS packages  
\usepackage{natbib}
\usepackage{hyperref}

\usepackage{longtable}
\usepackage{csvsimple} %Load CSV file as tables
\usepackage{adjustbox} %  Fit wide table

\usepackage{mathtools}
\usepackage{ifthen}

\usepackage{siunitx} % Scientific notation
\newcounter{biburlnumpenalty}
\newcounter{biburlucpenalty}
\newcounter{biburllcpenalty}

\def\){\right)} 
\def\({\left(} 
\def\]{\right]} 
\def\[{\left[}

\def\pBe{$^7\mathrm{Be}(p,\gamma)^8\mathrm{B}$}
\def\nLi{$^7\mathrm{Li}(n,\gamma)^8\mathrm{Li}$}

\def\nLiHead{$\bm{^7}$L\lowercase{i}$\bm{(n,\gamma)^8}$L\lowercase{i}}

\def\swaveHead{$\bm{^3S_1}$-$\bm{^3S_1^\star}$}
\def\pwaveHead{$\bm{^3P_2}$-$\bm{^3P_2^\star}$}

\def\EFTstarHead{EFT$_\star$}
\def\EFTgsHead{EFT$_\text{gs}$}

\def\P{\left(\frac{\stackrel{\rightarrow}{\nabla}}{m_c}
-\frac{\stackrel{\leftarrow}{\nabla}}{m_n}\right)}

\def\EFTg{EFT$_\text{gs}$}
\begin{document}

\title{Coupled-channel treatment of 
\texorpdfstring{\nLiHead}{n-Li7}
in effective field theory}

\author{%
Renato Higa}
\email{higa@if.usp.br}
\affiliation{Instituto de F\'isica, Universidade de S\~ao Paulo, R. do Mat\~ao Nr.1371, 05508-090, S\~ao Paulo, SP, Brazil}

\author{%
Pradeepa Premarathna}
\email{psp63@msstate.edu}
\affiliation{Department of Physics \& Astronomy and HPC$^2$ Center for 
Computational Sciences, Mississippi State
University, Mississippi State, MS 39762, USA}

\author{%
Gautam Rupak}
\email{grupak@ccs.msstate.edu}
\affiliation{Department of Physics \& Astronomy and HPC$^2$ Center for 
Computational Sciences, Mississippi State
University, Mississippi State, MS 39762, USA}

\begin{abstract}
The E1 contribution to the capture reaction \nLi ~is calculated at 
low energies.  We employ a coupled-channel formalism to account for the 
$^7\mathrm{Li}^\star$ excited core contribution. We develop a halo effective field theory power counting where capture in the spin $S=2$ channel is enhanced over the $S=1$ channel. A  next-to-leading order  calculation is presented where the excited core contribution is shown to affect only the overall normalization of the cross section. The momentum dependence of the capture cross section, as a consequence, is the same in a theory with or without the excited $^7\mathrm{Li}^\star$ degree of freedom at this order of the calculation. The kinematical signature of the $^7\mathrm{Li}^\star$ core is negligible at momenta below 1~MeV and significant only 
beyond the $3^+$ resonance energy, though still compatible with a 
next-to-next-to-leading order correction.
We compare our formalism with a previous  halo effective field theory calculation
[Zhang, Nollett, and Phillips, Phys. Rev. C {\bf 89}, 024613 (2014)] that also
treated the $^7\mathrm{Li}^\star$ core as an explicit degree of freedom. 
Our formal expressions and analysis disagree with this earlier work in several aspects.

\end{abstract}

%\date{\today}
\maketitle
 %==================================================
 \section{Introduction}
 \label{sec:Intro}
 %==================================================

Radiative capture reactions with light nuclei sharpen our knowledge about 
the primordial evolution of our universe, the fuel in the interior of 
stars, and explosive phenomena of astrophysical 
objects~\cite{Wiescher:2012aaa,Brune:2015nsv,Bertulani:2009mf}. 
Among them features the 
long-studied ${}^7{\rm Be}(p,\gamma)^8{\rm B}$ reaction, crucial in 
determining the flux of solar neutrinos that oscillate into different 
lepton flavors on their way to detection in Earth. 
The impact to neutrino oscillations and to the standard solar model 
of this reaction depends on the information of the respective 
cross section around the Gamow energy $\sim 20\,{\rm keV}$. 
The Coulomb repulsion at such low energies makes it extremely difficult, if 
not impossible, for direct measurements in laboratory, which are presently 
limited to as low as $\sim 100\,{\rm keV}$. 
Therefore, theoretical extrapolations of experimental 
data down to energies of astrophysical interest is an unavoidable necessity. 
The mirror-symmetry, backed by the accidental isospin symmetry of quantum 
chromodynamics (QCD) at low energies, is often invoked to constrain the 
strong 
nuclear part of the reaction. 
Besides the mirror connection, the ${}^7{\rm Li}(n,\gamma)^8{\rm Li}$ 
reaction may reveal its importance in some astrophysical scenarios, such as 
inhomogeneous Big-Bang nucleosynthesis or neutron-rich explosive 
environments~\cite{kawano:1991ApJ372,Wiescher:2012aaa}. 

Experimental studies on the ${}^7{\rm Li}(n,\gamma)^8{\rm Li}$ reaction date 
back to the 40's~\cite{PhysRev.72.646} followed by a handful of others between 
the 90's and 
2010's~\cite{Lynn:1991,Blackmon:1996,Heil:1998,Nagai:2005,Izsak:2013hga}. 
Due to the absence of the Coulomb repulsion, many measurements were done in 
the sub-keV region with good precision, thus serving as testbed for 
theoretical low-energy extrapolations. 
Theoretical descriptions of this reaction vary in degree of sophistication 
and accuracy --- \emph{ab initio}/microscopic 
models~\cite{Navratil:2010jn,Bennaceur:1999hv,Fossez:2015qxa,Descouvemont:1994gcx} 
are tied to the choice of two- and three-nucleon interactions and numerical 
precision of the method whereas 
two-body~\cite{Tombrello:1965,AURDAL1970385,PhysRevC.70.047603,Barker:2006a,PhysRevC.68.045802,Typel:2004us,Bertulani1996293,Huang:2008ye} 
and three-body~\cite{Shulgina:1996khz,Grigorenko:1999mp} 
cluster configuration with phenomenological potentials are much simpler 
and more flexible with adjustable parameters to fit data. Halo/cluster effective 
field theory (halo EFT) relies on the same cluster approach, but uses quantum 
field theory techniques and a small momentum ratio to provide a 
model-independent, systematic, and improvable expansion with controlled 
theoretical uncertainties. Halo EFT ideas emerge from 
Refs.~\cite{Bertulani:2002sz,Bedaque:2003wa} and were extended to several 
loosely-bound nuclear systems 
(see Refs.~\cite{Rupak:2016mmz,Hammer:2017tjm,Hammer:2019poc} and references therein). 

Halo EFT was first applied to ${}^7{\rm Li}(n,\gamma)^8{\rm Li}$ 
in Ref.~\cite{Rupak:2011nk}, which pointed out inconsistencies of 
potential-model extrapolations of fitted high-energy data to low energies, 
its origin, and ways to overcome it. The follow-up 
work~\cite{Fernando:2011ts} included E1 capture to the $1^{+}$ state of 
${}^8{\rm Li}$ and M1 capture from the initial $3^{+}$ resonant state, 
while the kinematical impact of the $\frac{1}{2}^-$ ${}^7{\rm Li}$ excited-core state 
was argued to be of higher-order. 
A subsequent work by Zhang, Nollett, and Phillips~\cite{Zhang:2013kja} 
combined halo EFT formalism with asymptotic normalization coefficients 
(ANCs) obtained from \emph{ab initio} variational Monte Carlo (VMC) calculation. 
Zhang {\em et al.} considered only E1 transitions, but included 
the $\frac{1}{2}^-$ excited state of ${}^7{\rm Li}$ as an explicit degree 
of freedom. Their overall results are qualitatively similar to ours within 
the leading-order (LO) theoretical uncertainties. They find the momentum-dependent 
contribution of the excited core of ${}^7{\rm Li}$ compatible with a 
next-to-leading order (NLO) correction, meeting the expectations of our 
initial assumptions~\cite{Fernando:2011ts}. 

Nevertheless, Zhang {\em et al.} raised critical comments to our work, 
namely, (a) that the excited state of ${}^7{\rm Li}$ must be formally 
included in a LO calculation of the $^8$Li bound state given its excitation energy of 
$\sim 0.5\,{\rm MeV}$, smaller than the $\sim 2\,{\rm MeV}$ neutron 
separation energy of ${}^8{\rm Li}$, (b) that the couplings of the 
different final state spin channels were made equal, which influences 
the rate of total initial spin  $S=2$ contribution to the capture 
reaction, and (c) that capture to the $1^{+}$ excited state of 
${}^8{\rm Li}$ was not a prediction, but an input to constrain our 
EFT parameters. Point (c) was our choice and true in certain sense, 
though the best we can do within halo EFT itself: 
in Ref.~\cite{Fernando:2011ts} we fixed two remaining renormalization constants 
(after fixing the usual ones from scattering lengths and binding energies) to 
just two input data, the thermal neutron capture cross sections to the ground 
$2^+$ and first excited $1^+$ states of ${}^8{\rm Li}$~\cite{Lynn:1991}, 
and postdict capture data at other energies. 
Zhang {\em et al.}, on the other hand, fix their remaining renormalization 
constants from their {\em ab initio} calculation. 
Answer to points (a) and (b) was the motivation of the present work. 

We include the $\frac{1}{2}^-$ excited state of ${}^7{\rm Li}$ core 
($^7\mathrm{Li}^\star$) as an explicit degree of freedom in a 
coupled-channel formalism, along similar lines of 
Refs.~\cite{Cohen:2004kf,Lensky:2011he}. In our formulation, the 
$^7\mathrm{Li}^\star$ excited core influences only the channel with 
total spin $S=1$, which is regarded as a NLO  contribution. 
At this order, the excited core contributions only affect the overall normalization of the capture cross section. The capture calculations show that  the kinematical impact of the excited core 
degree of freedom is negligible for momenta in the keV regime, and starts
being noticeable only beyond the $3^+$ resonant state of ${}^8{\rm Li}$, 
albeit compatible with a typical next-to-next-to-leading order (NNLO) correction. 
A power-counting is proposed that naturally incorporates the NLO contributions 
of the $S=1$ channel and $^7\mathrm{Li}^\star$ excited core, the latter 
with kinematic imprints only at NNLO. As a result, up to NLO the momentum dependence of the capture cross section in the EFT with explicit $^7\mathrm{Li}^\star$ degree of freedom is the same as that in the EFT from Refs.~\cite{Rupak:2011nk,Fernando:2011ts} that did not include the $^7\mathrm{Li}^\star$  explicitly. Once the overall normalization of the capture cross sections are fitted to data, the EFTs with and without excited $^7\mathrm{Li}^\star$ core give the same numerical result.

We also study the effect of the couplings, parameterized in terms of the effective momenta,  in the final $p$-wave state on capture rate and the spin $S=2$ branching ratio at thermal energy. 
 The use of the same value for the effective momenta in our previous works~\cite{Rupak:2011nk,Fernando:2011ts}, consistent with the power counting,  is shown to give satisfactory results well within a LO expectation. A NLO correction to this previous LO assumption is sufficient to bring both capture rate and branching ratio predictions in better agreement with data.

The paper is organized as follows. In Sec.~\ref{sec:interaction} we present 
the framework of halo EFT for two scenarios: 
without~\cite{Rupak:2011nk,Fernando:2011ts} and with the 
$^7\mathrm{Li}^\star$ excited core as an explicit degree of freedom, 
the latter with the pertinent channels that couple. 
Section~\ref{sec:Capture} contains the relevant formulas for the 
${}^7{\rm Li}(n,\gamma)^8{\rm Li}$ capture in both theories with 
and without $^7\mathrm{Li}^\star$. We also highlight the differences between our and Zhang {\em et al.}'s 
formulations for the $^7\mathrm{Li}^\star$ excited core in EFT. A survey of possible families of EFT 
parameters compatible with given observables is discussed in 
Sec.~\ref{sec:Survey}, followed by the proposed power-counting in 
Sec.~\ref{sec:powercounting}. Numerical results and analyses are presented 
in Sec.~\ref{sec:Results} with concluding remarks in 
Sec.~\ref{sec:Conclusions}. 

 %==================================================
 \section{Interaction}
 \label{sec:interaction}
 %==================================================

 In this section we construct two halo EFTs: one without the explicit $^7\mathrm{Li}^\star$ degree of freedom that we refer to as \EFTg ~and one with that we refer to as EFT$_\star$. In the former,  the low-energy degrees of freedom would be the spin-parity $\frac{1}{2}^+$ neutron, the $\frac{3}{2}^-$ ground state of $^7$Li, and the photon. In the latter we also have the $\frac{1}{2}^-$   excited   $^7\mathrm{Li}^\star$ state as an additional degree of freedom. The $2^+$ ground  and $1^+$ excited  states of $^8$Li are represented as $p$-wave bound states of the neutron and the $^7$Li ($^7\mathrm{Li}^\star$) core with binding momenta ${\gamma_0\approx\SI{57.78}{\mega\eV}}$ (${\gamma_\star\approx\SI{64.22}{\mega\eV}}$) and ${\gamma_1\approx\SI{41.56}{\mega\eV}}$ (${\gamma_{1\star}\approx\SI{50.12}{\mega\eV}}$), respectively~\cite{Tilley:2004}. 
 We identify the relative center-of-mass (c.m.) momentum $p$ and the binding momenta with a low momentum scale ${Q\sim\{\gamma_0,\gamma_1,\gamma_\star,\gamma_{1\star}\}\gtrsim p}$. The breakdown scale can be associated with pion physics and $^4$He-$^3$H break up of the core giving $\Lambda\sim 100-150$ MeV. Estimates of $\Lambda$ from the sizes of $p$-wave effective momenta suggest a larger breakdown scale $\Lambda\gtrsim 200$ MeV~\cite{Rupak:2011nk,Fernando:2011ts,Zhang:2013kja}. We assume a cutoff in between these two estimates and take the ratio $Q/\Lambda\sim 1/3$. We only consider the non-resonant E1 capture. The M1 contribution from the $3^+$ resonance  initial state has been considered in Ref.~\cite{Fernando:2011ts}, and would be revisited in a future publication~\cite{Higa:2020}.
 
 A radiative capture calculation requires a description of the initial scattering states, the final bound states, and the electroweak transition operators. The dominant contribution to capture at low momentum is from initial $s$-wave states to the final $p$-wave bound state through the E1 transition. The electroweak operators relevant to our calculation are one-body currents produced through minimal substitution by gauging the incoming core  momentum $\bm{q}\rightarrow\bm{q}+e Z_c\bm{A}$ where $Z_c=3$ is the charge of the $^7$Li/$^7\mathrm{Li}^\star$ core in units of the proton charge $e$. Thus we start with a
framework for 
the strong interaction that describes the initial scattering states, the final bound states, and the E1 operators. 
 
 The strong interaction in the first halo EFT (\EFTg) is given by the  Lagrangian
 %--------------- Equation-1 -----------------
 \begin{align}\label{eq:EFT}
\mathcal L={}& N^\dagger\left[i\partial_0+\frac{\nabla^2}{2 m_n}\right] N 
+C^\dagger\left[i\partial_0+\frac{\nabla^2}{2 m_c}\right]C
\nonumber\\
&+\sum_\zeta{\chi_{[j]}^{(\zeta)}}^\dagger\left[\Delta^{(\zeta)}+h^{(\zeta)}\left(i\partial_0
+\frac{\nabla^2}{2M}\right)\right]\chi_{[j]}^{(\zeta)}\nonumber \\
&+\sqrt{\frac{2\pi}{\mu}}\sum_\zeta
\left[
{\chi_{[j]}^{(\zeta)}}^\dagger N^T P_{[j]}^{(\zeta)}C+\operatorname{h.c.}
\right]\, ,
\end{align}
where $N$ represents the $\frac{1}{2}^{+}$ neutron with mass $m_n=939.6$ MeV, $C$ represents the $\frac{3}{2}^-$ $^7$Li core with mass 
$m_c=6535.4$ MeV~\cite{Huang:2017,Mang:2017},  $M=m_n+m_c$ is the total mass, and $\mu= m_n m_c/M$ is the reduced mass. We use natural units $\hbar=1=c$. 
The dimer fields  $\chi^{(\zeta)}_{[j]}$ are auxiliary that are introduced for convenience. They are particularly useful in resuming momentum dependent operators but it is equivalent to the original theory without the dimer~\cite{Bertulani:2002sz}. The summation in $\zeta$ is over the relevant $s$- and $p$-wave channels---$^3S_1$, $^5S_2$, $^3P_1$, $^3P_2$, $^5P_1$, $^5P_2$---written 
in the spectroscopic notation ${}^{2S+1}L_J$ with $S$ the total spin, $L$ the total orbital angular momentum, and $J$ the total angular momentum~\cite{Rupak:2011nk,Fernando:2011ts}. $P^{(\zeta)}_{[j]}$ are the projectors for a given channel $\zeta$ whose explicit forms are provided in Appendix~\ref{sec:Projectors}. The repeated subscript $[j]$ is summed over, and it is a single index or double indices as appropriate for $J=1$ and $J=2$ states, respectively.
For example, with $\zeta={}^3P_2$ one should read, using Eq.~(\ref{eq:projdefr-space}), $\chi^{(\zeta)}_{[j]}=\chi^{(^3P_2)}_{ij}$ and $P^{(\zeta)}_{[j]}= P^{(^3P_2)}_{ij}$. 

The neutron and $^7$Li core interaction, in Eq.~(\ref{eq:EFT}), 
is mediated through the exchange of the $\chi^{(\zeta)}_{[j]}$ fields.  These auxiliary fields can be integrated out of the theory to generate 
neutron-core contact interactions without changing the particle content, and therefore physical observables in the theory. For example, the $s$-wave amplitude calculated later in Eq.~(\ref{eq:swaveAmp}) is exactly the same as the one calculated in Refs.~\cite{Rupak:2011nk,Fernando:2011ts} that did not use an auxiliary dimer field for the $s$-wave interaction.
The unknown couplings $h^{(\zeta)}$ are included in the dimer propagator~\cite{Griesshammer:2004pe}, for convenience,  instead of an equivalent formulation where they appear in the dimer-particle interaction~\cite{Rupak:2011nk,Fernando:2011ts}. Again, this does not affect the calculation of physical observables. As an example, the capture cross section in this work is proportional to the product of the neutron-core-dimer coupling  $2\pi/\mu$ in Eq.~(\ref{eq:EFT}) and 
$\mathcal Z^{(^5P_2)}$ in Eq.~(\ref{eq:Zphi}) calculated from the $p$-wave amplitude. This is exactly the corresponding product of the neutron-core-dimer coupling  $h^{(^5P_2)}$ and $\mathcal Z^{(^5P_2)}$ in Eq.~(14) of Ref.~\cite{Fernando:2011ts}. 
The couplings $\Delta^{(\zeta)}$, $h^{(\zeta)}$ can in principle be related to elastic scattering phase shifts if available. The E1 contribution to \nLi ~has been calculated using this theory in Refs~\cite{Rupak:2011nk,Fernando:2011ts}. We will present the results in Section~\ref{sec:Capture}.
 
 The second halo EFT with excited $^7\mathrm{Li}^\star$ core (EFT$_\star$) can be described with the Lagrangian
 %--------------- Equation-2 -----------------
\begin{align}
 \label{eq:EFTstar}
\mathcal L_\star = N^\dagger\left[i\partial_0+\frac{\nabla^2}{2 m_n}\right] N 
 +C^\dagger\left[i\partial_0+\frac{\nabla^2}{2 m_c}\right]C
\nonumber \\
+C_\star^\dagger\left[i\partial_0-E_\star+\frac{\nabla^2}{2 m_c}\right]C_\star\nonumber \\
+\sum_{\zeta,\zeta'}{\chi_{[j]}^{(\zeta)}}^\dagger\left[
\Pi^{(\zeta\zeta')}+t^{(\zeta\zeta')}\left(i\partial_0
+\frac{\nabla^2}{2M}\right)\right]\chi_{[j]}^{(\zeta')}\nonumber\\
+\sqrt{\frac{2\pi}{\mu}}
\sum_{\zeta}\!\bigg[
{\chi_{[j]}^{(\zeta)}}^\dagger\! N^T\! P_{[j]}^{(\zeta)}\!C
\!+\!{\chi_{[j]}^{(\zeta)}}^\dagger\! N^T\! P_{[j]}^{(\zeta)}\!C_\star
\!+\!\operatorname{h.c.}
\bigg].
\end{align}
Here the $C_\star$ field represents the excited $^7\mathrm{Li}^\star$ core with excitation energy $E_\star=0.47761$ MeV~\cite{Tilley:2002}. There are a few other differences with the previous  \EFTg ~in Eq.~(\ref{eq:EFT}). In this theory, we have the 
additional scattering channels $^3S_1^\star$, $^3P_2^\star$, and $^3P_1^\star$ involving the $C_\star$ field. We also allow for the possibility of mixing between 
channels which is induced by the off-diagonal terms in the dimer field $\chi^{(\zeta)}_{[j]}$ (inverse) propagator matrix. The $^1P_1^\star$ channel does not contribute to the capture through E1 transition but can be included in the coupled-channel calculation of the $p$-wave $1^+$ bound state. The couplings $\Pi^{(\zeta\zeta')}$ and $t^{(\zeta\zeta')}$ can  be related to scattering phase shifts if available. 
The generic index $\zeta$ is used to represent all the channels, and one has to appropriately consider interactions only in the relevant channels as discussed below. 
A low-momentum scale $\gamma_\Delta=\sqrt{2\mu E_\star}\approx\SI{28.0}{\mega\eV}\sim Q$ is associated with the excited core~\cite{Zhang:2013kja}.

We present  coupled-channel calculations for mixing in $^3S_1$-$^3S_1^\star$ and $^3P_2$-$^3P_2^\star$ below. However, the formalism can be extended to other scattering channels as well. One could do the same to include mixing between all $S=1$ and $S=2$ $p$-wave channels. 
We discuss later in subsection \ref{sec:PwaveMix} our particular choice of $p$-wave mixing and the consequence of alternatively mixing all the possible $p$-wave channels 
such as $^5P_1$-$^3P_1$-$^3P_1^\star$-$^1P_1^\star$ of the $1^+$ excited state.

%======================================================
\subsection{\texorpdfstring{\swaveHead}{3S1-3S1*} Coupled-Channel}
\label{sec:SwaveMix}
%==================================================

A coupled-channel calculation involving  $s$-wave states was presented in Ref.~\cite{Cohen:2004kf}. See Ref.~\cite{Lensky:2011he} as well for a coupled-channel calculation with and without Coulomb interaction for $s$-wave bound states. Here we present a slightly different derivation using the dimer fields instead of 
nucleon-core contact interactions. 
Some of the renormalization conditions are a little different but the final results expressed in terms of scattering parameters are equivalent. The use of auxiliary fields  does not change physical observables in quantum field theory. We demonstrate this with explicit calculations.

The coupled-channel $s$-wave scattering amplitude is a $2\times2$ matrix written as
%--------------- Equation-3 -----------------
\begin{align}
    i \mathcal A^{(ab)}(p) =-\frac{2\pi}{\mu} 
i\mathcal D^{(ab)}(E,0)\, ,
\end{align}
where $E=p^2/(2\mu)$ is the c.m. energy and the superscripts are the row-column indices of the amplitude matrix. We identify the $^3S_1$ state as channel 1, and the 
$^3S_1^{\star}$ 
state as channel 2. The dimer propagator is given by 
%--------------- Equation-4 -----------------
\begin{align}
    \mathcal D(E,0) = \mathcal D_0(E,0)+
    \mathcal D_0(E,0)\Sigma(E,0) \mathcal D(E,0)\, ,
\end{align}
which is conveniently calculated from its inverse
%--------------- Equation-5 -----------------
\begin{align}
    \mathcal D^{-1}= \mathcal D_0^{-1} -\Sigma\, ,
\end{align}
where $\mathcal D_0^{-1}$ is the inverse free dimer propagator and $\Sigma$ is the self-energy. We have the free inverse dimer propagator 
directly 
from Eq.~(\ref{eq:EFTstar}):
%--------------- Equation-6 -----------------
\begin{align}
    [\mathcal D_0(E,0)]^{-1}&= 
\begin{pmatrix}
\Pi^{(11)}   & 
    \Pi^{(12)} \\
  \Pi^{(12)} & \Pi^{(22)}
\end{pmatrix}
\, ,
\end{align}
where we only kept the couplings $\Pi^{(ij)}$ in a low-momentum expansion. In a single-channel calculation this would correspond to keeping only the scattering length contribution. The self-energy loop integral is 
%--------------- Equation-7 -----------------
\begin{align}\label{eq:swaveSigma}
   - \Sigma(E,0)
&=-\frac{2\pi}{\mu}   
\begin{pmatrix}
J_0(-ip) & 0\\
  0& J_0(-ip_{\star})
\end{pmatrix}\,,\nonumber\\[3mm]
    J_0(x)&=-2\mu\left(\frac{\lambda}{2}\right)^{4-D}
    \int\frac{d^{D-1}\bm q}{(2\pi)^{D-1}}\frac{1}{q^2+x^2}
\nonumber\\
    &=-\frac{\mu}{2\pi}(\lambda-x)\, ,
\end{align}
where $p_{\star}=\sqrt{p^2-\gamma_{\Delta}^2+i0^+}$
and $\lambda$ is the renormalization group (RG) scale. We use dimensional regularization in the so-called power divergence subtraction (PDS) scheme~\cite{Kaplan:1998tg} that removes all divergences in space-time dimensions $D\leq 4$. The scattering amplitude has to be independent of $\lambda$ which can be accomplished with the renormalized couplings
%--------------- Equation-8 -----------------
\begin{align}\label{eq:swaveRG}
    \Pi^{(ij)}=\frac{1}{a_{ij}}-\lambda\delta_{ij}\, ,
\end{align}
where we introduced the scattering lengths $a_{ij}$ following Ref.~\cite{Cohen:2004kf}.
Our RG condition differs in the overall sign of $a_{12}$.
We get 
%--------------- Equation-9 -----------------
\begin{align}
    [\mathcal D(E,0)]^{-1}= 
\begin{pmatrix}
\frac{1}{a_{11}}  +i p & {} &
  \frac{1}{a_{12}} \\
   {} & {}\\
  \frac{1}{a_{12}} & {} &
  \frac{1}{a_{22}} +ip_{\star}
\end{pmatrix}\, ,
\end{align}
and in particular the $s$-wave amplitude
%--------------- Equation-10 -----------------
\begin{align}
    \mathcal A^{(11)}(p)&=\frac{2\pi}{\mu}\frac{-\frac{1}{a_{22}}
    -ip_{\star}}{
    (-\frac{1}{a_{11}}-i p)
    (-\frac{1}{a_{22}}-ip_{\star})
    -\frac{1}{a_{12}^2}}\, .
\end{align}
The off-diagonal amplitude mixing channels 1 and 2 is
%--------------- Equation-11 -----------------
\begin{align}
    \mathcal A^{(12)}(p) &= \frac{2\pi}{\mu}\frac{1/a_{12}}{
    (-\frac{1}{a_{11}}-i p)
    (-\frac{1}{a_{22}}-ip_{\star})
    -\frac{1}{a_{12}^2}}\, .
\end{align}
These expressions agree with Eq.~(2.19) from Ref.~\cite{Cohen:2004kf} in a theory without auxiliary dimer fields, except the sign of $a_{12}$. When the coupling $\Pi_{12}=0$, there is no mixing in the EFT$_\star$ between the two channels, and the coupled-channel calculation reduces to two single-channel calculations as expected.

At momenta $p\ll\gamma_\Delta$, the scattering amplitudes are analytic around $p=0$, and in particular $\mathcal A^{(11)}$ is given by the effective range expansion (ERE). So we need
%--------------- Equation-12 -----------------
\begin{align}
&-\frac{1}{a_{11}}-a_{12}^{-2}/(-\frac{1}{a_{22}}+\sqrt{-p^2+\gamma_\Delta^2-i0^+})\nonumber\\
&\approx 
     -\frac{1}{a_{11}}+\frac{a_{22}a_{12}^{-2}}{1-a_{22}\gamma_\Delta}-
 \frac{1}{2}\frac{a_{22}^2 a_{12}^{-2}}{\gamma_\Delta(1-a_{22}\gamma_\Delta)^2}p^2
+\dots\nonumber \\
&=-\frac{1}{a_0^{(1)}}+\frac{1}{2}r_0^{(1)} p^2+\dots, 
\end{align}
where $a_0^{(1)}$ and $r_0^{(1)}$ are the scattering length and effective range in the $^3S_1$ channel, respectively. Matching the EFT expression to the ERE one obtains 
%--------------- Equation-13 -----------------
\begin{align}
    a_{11} &= a_0^{(1)}\frac{1-a_{22}\gamma_\Delta}
    {1+a_{22}(a_0^{(1)} a_{12}^{-2}-\gamma_\Delta)}\, ,  \nonumber \\[3mm] 
    a_{12}^{-2} &= - r_0^{(1)}\gamma_\Delta \frac{(1-a_{22}\gamma_\Delta)^2}{a_{22}^2}\, .
\label{eq:EFTstarConstrains}
\end{align}
We can fix $a_{11}$ from $a_0^{(1)}$ and  $a_{12}$ (up to a sign) from $r_0^{(1)}$,  leaving $a_{22}$ as an undetermined parameter. In principle $a_{22}$, $a_{12}$ can be obtained from the low-momentum measurement of $\mathcal A^{(12)}$~\cite{Knox:1981dgp}. One should note that in EFT$_\star$, an effective range $r_0^{(1)}$ is generated dynamically though we started with a momentum-independent interaction because there is a momentum associated with the excitation energy of the core. Ref.~\cite{Cohen:2004kf} considered the situation where the scattering lengths $a_{ij}\sim 1/Q$ are fine-tuned. 
In the $^8$Li system $a_0^{(1)}=\SI{0.87+-0.07}{\femto\meter}$~\cite{Koester:1983} which is of natural size $1/\Lambda$. Thus we assume that all the scattering lengths $a_{ij}\sim 1/\Lambda$ are of natural size.
From the last relation in Eq.~(\ref{eq:EFTstarConstrains}) we see that $r_0^{(1)}$ has to be negative. Further, one notices that $1-a_{22}\gamma_\Delta\sim 1$ for  $a_{22}\sim 1/\Lambda$ which gives a fine-tuned $|r_0^{(1)}|\sim a_{22}^2 a_{12}^{-2}\gamma_\Delta^{-1}\sim 1/Q$ for $a_{22}\sim1/\Lambda\sim a_{12}$, $\gamma_\Delta\sim Q$.
We can expand the scattering amplitudes in the $Q/\Lambda$ ratio and write
%--------------- Equation-14 -----------------
\begin{align}
\mathcal A^{(11)}(p) \approx{}&-\frac{2\pi}{\mu}a_0^{(1)}\left[
    1-i a_0^{(1)} p\right.\nonumber\\
    &\left.- i a_0^{(1)} {a_{22}^2}{ a_{12}^{-2}} 
    (p_\star-i\gamma_\Delta)+\dots\right]\, ,\nonumber\\
  \mathcal A^{(12)}(p) \approx{}&\frac{2\pi}{\mu}a_0^{(1)}\frac{a_{22}}{a_{12}}\left[
   1-i a_0^{(1)} p-i a_{22} p_\star\right.\nonumber\\
   &\left.- i a_0^{(1)} {a_{22}^2}{ a_{12}^{-2}} (p_\star-i\gamma_\Delta)
   +\dots\right]\, .
   \label{eq:EFTexpansionSwave}
\end{align}
The $Q/\Lambda$ expansion in Eq.~(\ref{eq:EFTexpansionSwave}) implies that the non-analyticity from the open channel involving the excited core $^7\mathrm{Li}^\star$ is a subleading effect. We postpone the comparison to the previous calculation by Zhang \emph{et al.}~\cite{Zhang:2013kja} to subsection~\ref{sec:Zhang1} to explain the impact on the total cross section instead of just the $s$-wave elastic scattering.

%======================================================
\subsection{\texorpdfstring{\pwaveHead}{3P2-3P2*} Coupled-Channel}
\label{sec:PwaveMix}
%==================================================

The coupled-channel calculation for $p$-wave states is very similar to the $s$-wave states in subsection~\ref{sec:SwaveMix}. An important difference is that for $p$ waves we need effective momentum corrections at LO~\cite{Bertulani:2002sz,Bedaque:2003wa}. Now we write the free inverse dimer propagator as
%--------------- Equation-15 -----------------
\begin{align}
    [\mathcal D_0(E,0)]^{-1}&= 
\begin{pmatrix}
\Pi^{(11)} +  E t^{(11)} & 
    \Pi^{(12)} +  E t^{(12)} \\ %%{}&{}\\
  \Pi^{(12)} + E t^{(12)} &  \Pi^{(22)} +  E t^{(22)}
\end{pmatrix}\, ,
\end{align}
where we identify $^3P_2$ as channel 1 and $^3P_2^\star$ as channel 2. The $p$-wave self-energy term is given by
%--------------- Equation-16 -----------------
\begin{align}
   - \Sigma(E,0)&=-\frac{6\pi}{\mu^3}
\begin{pmatrix}
J_1(-ip)  & 0\\
  0& J_1(-ip_{\star}) 
\end{pmatrix}\, ,\nonumber \\[3mm]
    J_1(x)&=-\frac{2\mu}{D-1}\left(\frac{\lambda}{2}\right)^{4-D}
    \int\frac{d^{D-1}\bm q}{(2\pi)^{D-1}}\frac{q^2}{q^2+x^2}
\nonumber\\
&=-\frac{\mu}{6\pi}\left( x^3-\frac{3}{2}x^2\lambda  +\frac{\pi}{2}\lambda^3\right)\, .
\end{align}
The RG conditions (no sum over repeated indices intended)
%--------------- Equation-17 -----------------
\begin{align}\label{eq:pwaveRG}
   \mu^2 \Pi^{(ij)}&= \frac{1}{a_{ij}}-\frac{\pi}{2}\lambda^3\delta_{ij}+\frac{3}{2}\gamma_\Delta^2\lambda \delta_{i2}\delta_{ij}\, , \nonumber\\
    \mu t^{(ij)}&= -r_{ij} - 3\lambda\delta_{ij}\, ,
\end{align}
make the inverse propagator (and the scattering amplitude) $\lambda$-independent:
%--------------- Equation-18 -----------------
\begin{align}
&   [\mathcal D(E,0)]^{-1}= \nonumber\\[3mm]
&\frac{1}{\mu^2}
\begin{pmatrix}
\frac{1}{a_{11}} -\frac{1}{2} r_{11} p^2 +i p^3 & 
  \frac{1}{a_{12}} -  \frac{1}{2} r_{12}p^2\\
  {} & {} \\
  \frac{1}{a_{12}} -  \frac{1}{2} r_{12}p^2& 
  \frac{1}{a_{22}} -\frac{1}{2} r_{22} p^2 +ip_{\star}^3
\end{pmatrix}\, ,
\end{align}
where $a_{ij}$ are the $p$-wave scattering volumes, and $r_{ij}$ that carry units of momentum are the $p$-wave effective momenta. We use the same notation for the scattering volume as used for the scattering length in the previous subsection. However, the scattering volumes do not contribute to the cross section, and only appear here. Thus it should not cause any confusion. The $p$-wave coupled-channel amplitude is given by 
%--------------- Equation-19 -----------------
\begin{align}
\mathcal A(p)=
-\frac{2 \pi}{ \mu} \frac{p^2}{\mu^2} \mathcal D(E,0)\, .
\end{align}
The coupled-channel amplitude is expected to have a single pole at positive imaginary momentum $p=+i\gamma_0$ (or $p_\star=+i\sqrt{\gamma_0^2+\gamma_\Delta^2}\equiv +i\gamma_\star$) 
associated with  the $^8$Li bound state. This can be made more explicit when specifying the  RG conditions from the ERE around the binding momentum in Eq.~(\ref{eq:pwaveRG}) as~\cite{Fernando:2011ts}
%--------------- Equation-20 -----------------
\begin{align}
   a_{11}^{-1} &=-\gamma_0^3 -\frac{1}{2} r_{11} \gamma_0^2\, , & 
   a_{22}^{-1} &=-\gamma^3_\star -\frac{1}{2} r_{22}\gamma_0^2\, ,\nonumber\\
   a_{12}^{-1} &= -\frac{r_{12}}{2}\gamma_0^2\, .
\end{align}
This gives for momentum $p\approx i\gamma_0$ ($p_\star\approx i\gamma_\star$),
%--------------- Equation-21 -----------------
\begin{align}\label{eq:poleExp}
 [\mathcal D(E,0)]^{-1}   \approx-i\frac{\gamma_0(p-\!i\gamma_0)}{\mu^2}\!\begin{pmatrix}
r_{11}+\!3\gamma_0  &r_{12} \\
   r_{12} & r_{22} +\!3\gamma_\star
\end{pmatrix}\nonumber\\
=-\frac{E+\!B_0+\!i0^+}{\mu}\!\begin{pmatrix}
r_{11}+\!3\gamma_0  &r_{12} \\
   r_{12} & r_{22} +\!3\gamma_\star
\end{pmatrix}\, ,
\end{align}
generating a single pole in $\mathcal D(E,0)$ at negative energy $E=-B_0=-\gamma_0^2/(2\mu)$ on the first energy Riemann sheet corresponding to a bound state. We verified numerically that there are no other shallow bound 
or resonance states for typical $r_{ij}\sim\Lambda$ and experimental ANCs. The residue at the energy pole of the amplitude is related to the wave function renormalization constant which is usually calculated directly as 
%--------------- Equation-22 -----------------
\begin{align}
    [\mathcal Z]^{-1} &= \frac{d}{d E} [\mathcal D(E,0)]^{-1}\Big|_{E=-B_0}\nonumber\\
    &= 
    - \frac{1}{\mu}
\begin{pmatrix}
r_{11}+3\gamma_0  &r_{12} \\
   r_{12} & r_{22} +3\gamma_\star
\end{pmatrix}\, .
\label{eq:ZphiMix}
\end{align}
The wave function renormalization constant reduces to the single-channel result~\cite{Rupak:2011nk,Fernando:2011ts} when $r_{12}$ vanishes. In our calculation we will not assume $r_{12}$ to be small. We will simply fit the constants $\mathcal Z_{11}$,  $\mathcal Z_{22}$  to the ANCs without attempting to interpret what it implies in terms of the scattering parameters because the two $^3P_2$, $^3P^\star_2$ ANCs are not sufficient to determine the three effective momenta $r_{ij}$.

We briefly explore the consequences of including more $p$-wave channels in the mixing, such as the  $^5P_2$-$^3P_2$-$^3P^\star_2$ case. The  calculation would be similar to the $^3P_2$-$^3P^\star_2$ case with a diagonal self-energy $\Sigma(E,0)$ matrix. The inverse free dimer propagator $[\mathcal D_0(E,0)]^{-1}$ containing off-diagonal terms would depend on 6 scattering volumes $a_{ij}$ and 6 effective momenta $r_{ij}$. The $\mathcal Z_{ij}$s would depend on a combination of $r_{ij}$s. Though the expressions for $\mathcal Z_{ij}$ in a three coupled-channel calculation in terms of effective momenta would differ from the two coupled-channel calculation, the numerical values obtained by fitting the $\mathcal Z_{ij}$s to the capture data (or ANCs) would remain the same and  would not affect the momentum dependence of the capture calculation. The single-channel treatment of $^5P_2$ is compatible with the fact that the $\mathcal Z$  in this channel is much larger than the other two $p$-wave channels, and that the measured $p$-wave phase shifts for $^7\mathrm{Li}(n,n)^7\mathrm{Li}$ and $^7\mathrm{Li}(n,n')^7\mathrm{Li}^\star$ are of different sizes~\cite{Knox:1981dgp}.

In the initial $s$-wave scattering, the $S=2$ and $S=1$ spin channels do not mix as they carry different total angular momentum.  Given that the one-body E1 currents cannot change spins and the $^5P_2$ channel is treated separately from the $S=1$ $p$-wave channels, the capture in the spin $S=2$ and $S=1$ channels do not mix, and they are calculated separately.  

A  $^5P_1$-$^3P_1$-$^3P^\star_1$-$^1P_1\star$ coupled channel calculation is straightforward. However, we follow the formalism for the $2^+$ ground state calculation, for simplicity, and treat $^5P_1$ as a single-channel.
 The spin $S=2$ and $S=1$ initial $s$-wave states are treated separately as discussed earlier. The one-body E1 currents do not change the initial spin state. 
Thus, the capture in the $S=2$ and $S=1$ channels can be calculated separately when $^5P_1$ is treated as a single-channel.
 A $^3P_1$-$^3P^\star_1$-$^1P_1\star$ coupled-channel calculation  would result in wave function renormalization constants $\mathcal Z_{ij}$s that depend on 6 effective momenta. We determine these from the capture data (or ANCs) without attempting to interpret them in terms of the short distance physics contained in the effective momenta.

From the analysis above, we can see that, in the final state, channel mixing is not crucial in the capture calculation and gives the same numerical result as long as we  fit the wave function renormalization constants $\mathcal Z_{ij}$ to capture data (or ANCs)  without attempting to interpret their dependence on the $p$-wave effective momenta. 
Channel mixing in the initial state is needed, however, to include the kinematical dependence of the capture cross section on the excited $^7\textnormal{Li}^\star$ core degree of freedom.

We postpone the comparison to the $p$-wave bound state calculation by Zhang \emph{et al.}~\cite{Zhang:2013kja} to subsection~\ref{sec:Zhang1}.

 %==================================================
 \section{Capture calculation}
 \label{sec:Capture}
 %==================================================
 
\begin{figure}[tbh]
\begin{center}
\includegraphics[width=0.4\textwidth,clip=true]{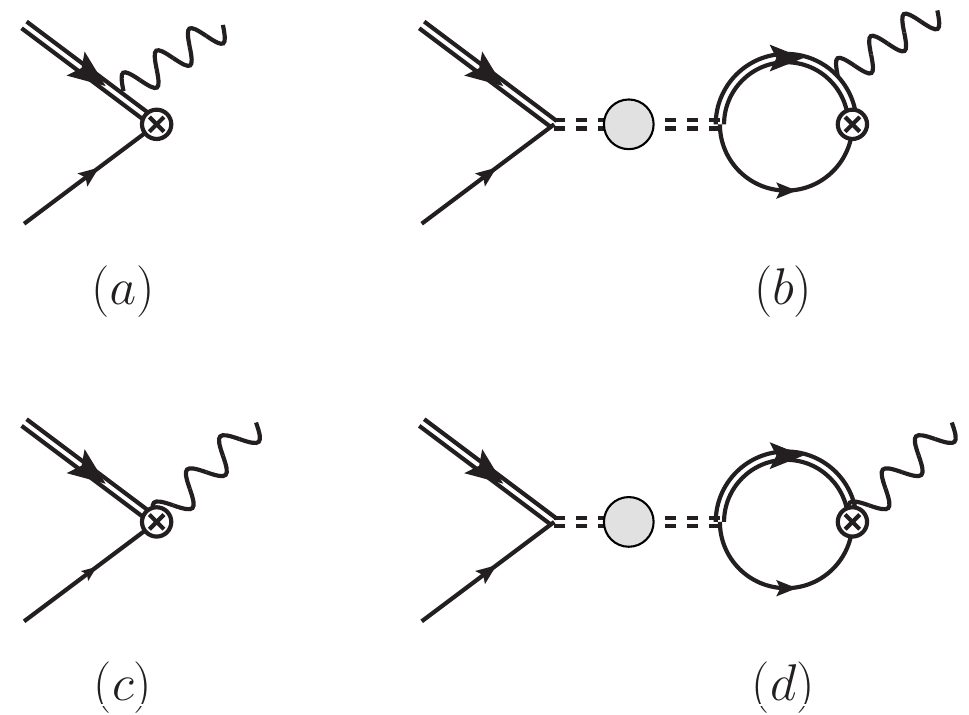}
\end{center}
\caption{\protect Double solid line represents a $^7$Li or $^7\mathrm{Li}^\star$ core as appropriate, single solid line a neutron, double dashed line a dressed dimer propagator, wavy line a photon, and $\otimes$ represents the final bound state.}
\label{fig:Capture}
\end{figure}

The E1 capture reaction \nLi ~is given by the diagrams in Fig.~\ref{fig:Capture}. 
The excited core $^7\mathrm{Li}^\star$ contributes only in the $S=1$ channel. Thus the $S=2$ channel calculation in both  \EFTg ~and EFT$_\star$ are very similar. 
 %%------------- Figure-1 --------------------------------

We start with the calculation in the $S=2$ channel in \EFTg ~first. The squared amplitude for the capture from the initial $^5S_2$ state to the $2^+$ ground state $^8$Li is~\cite{Rupak:2011nk,Fernando:2011ts} 
 %--------------- Equation-23 -----------------
\begin{align}
  \left|\mathcal M^{(^5P_2)}_\mathrm{E1}\right|^2
  ={}& (2J+1) \left(\frac{Z_c m_n}{M}\right)^2\frac{64\pi\alpha M^2}{\mu}
    \frac{2\pi}{\mu}\mathcal Z^{(^5P_2)}\nonumber\\
    &\times\bigg|1-\frac{2}{3}\frac{p^2}{p^2+\gamma_0^2}
\nonumber\\
   &-\mathcal A_0(a_0^{(2)}, p)
     [B_0(p,\gamma_0)+J_0(-ip)]\bigg|^2\, ,
\label{eq:E1amplitude}
 \end{align}
 where $J=2$ and $\alpha=e^2/(4\pi)=1/137$. The initial state strong interaction in the $s$-wave is given by the amplitude
 %--------------- Equation-24 -----------------
 \begin{align}\label{eq:swaveAmp}
     \mathcal A_0(a_0^{(2)},p)=\frac{2\pi}{\mu}\frac{1}{-\frac{1}{a_0^{(2)}}-ip}\, ,
 \end{align}
 with the scattering length 
$a_0^{(2)}=\SI{-3.63+-0.05}{\femto\meter}$~\cite{Koester:1983} 
scaling as $1/Q$. The use of auxiliary dimer field $\chi^{(^5S_2)}$ in deriving  Eq.~(\ref{eq:swaveAmp}) results in the same expression as the calculation without auxiliary field in Refs.~\cite{Rupak:2011nk,Fernando:2011ts}.
The loop contribution from 
diagram $(b)$ of Fig.~\ref{fig:Capture} is contained in the function 
 %--------------- Equation-25 -----------------
 \begin{align}
     B_0(p,\gamma)&=\left(\frac{\lambda}{2}\right)^{4-D}
     \int\frac{d^{D-1}\bm q}{(2\pi)^{D\!-\!1}}
     \frac{q^2\;4\mu/(D-1)}{(q^2\!-\!p^2\!-\!i0^+)(q^2\!+\!\gamma^2)}\nonumber\\
     &=\mu\left(
     \frac{\lambda}{2\pi}+\frac{1}{3\pi}\frac{ip^3-\gamma^3}{p^2+\gamma^2}
     \right)\, .
 \end{align}
 Fig.~\ref{fig:Capture} $(d)$ is proportional to $J_0(-ip)$. 
 The combination $B_0(p,\gamma)+J_0(-i p)$ is RG scale independent.
The wave function renormalization constant in the $^5P_2$ channel, assuming a single-channel calculation, has the simple form
 %--------------- Equation-26 -----------------
 \begin{align}\label{eq:Zphi}
     \frac{2\pi}{\mu}\mathcal Z^{(^5P_2)}=-\frac{2\pi}{r_1^{(^5P_2)}+3\gamma_0}\, ,
 \end{align}
 where $r_1^{(^5P_2)}$ is the $p$-wave effective momentum in the $^5P_2$ channel.
 In \EFTg, the capture from initial $^3S_1$ state to the $2^+$ ground state is given by a similar expression as above in Eq.~(\ref{eq:E1amplitude}) with the replacements $a_0^{(2)}\rightarrow a_0^{(1)}$ for the scattering length in the $^3S_1$ channel, and $r_1^{(^5P_2)}\rightarrow r_1^{(^3P_2)}$ for the effective momentum in the $^3P_2$ channel. 
 
 To calculate the cross section, we need to combine the contributions from the spin $S=2$ and $S=1$ channels with appropriate weights. We use the result of a three nuclear-cluster microscopic treatment from Ref.~\cite{Grigorenko:1999mp} that finds the $2^+$ ground state of $^8$Li is composed mostly of a $p_{3/2}$ valence neutron state. 
  The Clebsch-Gordan decomposition of the $p_{3/2}$ state in terms of the $^5P_2$ and $^3P_2$ state for the total angular momentum-parity $J^\pi=2^+$, azimuthal number $J_z=m$  is then  
 %--------------- Equation-27 -----------------
 \begin{multline}\label{eq:2plusstate}
     |2^+, m\rangle= a|S=2, L=1, J=2,J_z=m\rangle \\
     +\sqrt{1-|a|^2}\,|S=1, L=1, J=2, J_z=m\rangle\, ,
 \end{multline}
 with $a=1/\sqrt{2}$ in the standard Condon and Shortley sign convention. The $p_{3/2}$ configuration implies equal contributions from the  $^5P_2$ and $^3P_2$ states to the $2^+$ $^8$Li ground state.
 The total c.m. cross section for the capture to the $2^+$ state in \EFTg~is, then, given by
 %--------------- Equation-28 -----------------
 \begin{align}\label{eq:crosssection}
 \sigma_\mathrm{E1}^{(2^+)}(p)={}&\frac{1}{16\pi M^2}\frac{k_0}{p}\frac{1}{8}
\bigg[ |a|^2 \left|\mathcal M_\mathrm{E1}^{(^5P_2)}\right|^2
\nonumber\\
&+(1-|a|^2)\left|\mathcal M_\mathrm{E1}^{(^3P_2)}\right|^2\bigg]\, .
 \end{align}

It is straightforward to extend this calculation for the capture to the $1^+$ excited state---one replaces
$J=1$ in Eq.~(\ref{eq:E1amplitude}), $\gamma_0\rightarrow\gamma_1$ for the $1^+$ state binding momentum, and uses the appropriate $p$-wave effective momenta in the $^3P_1$ and $^5P_1$ final states. To determine the relative contributions of the $^5P_1$ and $^3P_1$ states to the $1^+$ excited state of $^8$Li, we again use Ref.~\cite{Grigorenko:1999mp} that finds $1^+$ to be mostly composed of the $p_{1/2}$ neutron state. The Clebsch-Gordan decomposition of $p_{1/2}$ in terms of the $^5P_1$ and $^3P_1$ states reads
%--------------- Equation-29 -----------------
 \begin{align}\label{eq:1plusstate}
     |1^+,m\rangle={}&b|S=2, L=1, J=1,m\rangle
\nonumber\\&
-\sqrt{1-|b|^2}\,|S=1,L=1,J=1,m\rangle\, ,
 \end{align}
with $b=\sqrt{5/6}$ in the standard sign convention. The choice $b=1/\sqrt{2}$ in Ref.~\cite{Fernando:2011ts} was an error that we correct here, see Fig.~\ref{fig:TotalCaptureANC}.

 Next we discuss the capture process in EFT$_\star$. The capture in the spin $S=2$ remains exactly the same as Eq.~(\ref{eq:E1amplitude}). The additional contributions in EFT$_\star$ come from Figs.~\ref{fig:Capture} $(b)$ and $(d)$ in the spin $S=1$
where the initial state $s$-wave interactions have to be described in terms of coupled-channel amplitudes $\mathcal A^{(11)}$ and $\mathcal A^{(12)}$ derived earlier in 
 Eq.~(\ref{eq:EFTexpansionSwave}). The contribution from $\mathcal A^{(12)}$ also entails modifying the momentum of the core in the loops with a photon attached in Fig.~\ref{fig:Capture} $(b)$ and $(d)$. The spin $S=1$ $p$-wave contribution to $^8$Li is included as a coupled-channel calculation as discussed in subsection~\ref{sec:PwaveMix} with the relative contributions of the channels determined by the coupled-channel $\mathcal Z^{(^3P_2)}$, $\mathcal Z^{(^3P_2^\star)}$. 
A direct calculation of capture to the $2^+$ ground state in spin $S=1$ channel is given by
 %--------------- Equation-30 -----------------
\begin{align}
\left|\mathcal M^{(^3P_2)}_\mathrm{E1,\star}\right|^2
&=(2J+1)\left(\frac{Z_c m_n}{M}\right)^2\frac{64\pi\alpha M^2}{\mu}
\frac{2\pi}{\mu}\mathcal Z^{(^3P_2)}
\nonumber\\
&\times\bigg|1\!-\!\frac{2}{3}\frac{p^2}{p^2\!+\!\gamma_0^2}
-\!\mathcal A^{(11)}( p)\left[B_0(p,\gamma_0)\!+\!J_0(-ip)\right]
\nonumber\\ 
&-\mathcal A^{(12)}(p)\left[
B_0\left(p_{\star},\sqrt{\gamma_0^2+\gamma_\Delta^2}\right)
+J_0\left(-ip_{\star}\right)\right]
\nonumber\\ 
&\times
\frac{\sqrt{\mathcal Z^{(^3P_2^\star)}}}{\sqrt{\mathcal Z^{(^3P_2)}}}
\bigg|^2\, .
\label{eq:E1amplitudeStar}
 \end{align}
 
 In EFT$_\star$  we get 
 %--------------- Equation-31 -----------------
 \begin{align}\label{eq:corsssectionStar}
\sigma_\mathrm{E1,\star}^{(2^+)}(p)={}&\frac{1}{16\pi M^2}\frac{k_0}{p}\frac{1}{8}
\bigg[ |a|^2\left|\mathcal M^{(^5P_2)}_\mathrm{E1}\right|^2
\nonumber\\&
+(1-|a|^2)\left|\mathcal M^{(^3P_2)}_\mathrm{E1,\star}\right|^2\bigg]\, ,
 \end{align}
 with $|a|^2=1/2$. The relative contribution to capture in the spin $S=2$ channel, which is not treated as a coupled-channel, remains the same in \EFTg ~and EFT$_\star$.
 
 In EFT$_\star$, the extension  to the capture to the $1^+$ state with 
 $J=1$ is straightforward 
with  $\gamma_0\rightarrow \gamma_1$, $|a|^2\rightarrow |b|^2=5/6$, and the appropriate modification of the wave function renormalization 
constants $\mathcal Z^{(^3P_2)}\rightarrow\mathcal Z^{(^3P_1)}$, $\mathcal Z^{(^3P_2^\star)}\rightarrow\mathcal Z^{(^3P_1^\star)}$ for the $^3P_1$ and $^3P_1^\star$ states, respectively. 
 
  The final state decomposition used in the \EFTg ~and EFT$_\star$ in terms of the $S=2,1$ channels gives an accurate description of the thermal~\cite{Lynn:1991} and sub-thermal~\cite{Heil:1998}   cross section. In Fig.~\ref{fig:TotalCaptureANC} and the $S=2$ branching ratio (bottom 2 rows) in Tables \ref{table:Li8Table} and \ref{table:Li8TableB}, the \EFTg/EFT$_\star$ predictions use only the measured ANCs~\cite{Trache:2003ir} and $a_0^{(2)}$ as input.
 
 %======================================================
\subsection{Previous mixed-channel calculation}
\label{sec:Zhang1}
%=======================================================

In this subsection we compare our formalism with the previous work by Zhang \emph{et al.}~\cite{Zhang:2013kja}. Their calculation differs from our result for the $^7\mathrm{Li}^\star$ contribution primarily in the treatment of the $p$-wave bound state channels, the incoming $s$-wave channels, and the power counting. We will comment on the differences in the latter after we propose our power counting in section~\ref{sec:powercounting}.  

In the final $p$-wave $2^+$ ground state calculation, the interaction used by Zhang \emph{et al.}~\cite{Zhang:2013kja} produces mixing between all $p$-wave channels:  $^5P_2$, $^3P_2$ and  $^3P_2^\star$, see Eq.~(14) in Ref.~\cite{Zhang:2013kja} whereas we only mix the $^3P_2$ and  $^3P_2^\star$ channels. The difference in the number of  mixed channels is not crucial to the analysis but how the mixing is implemented. 
The three-channel mixing in Ref.~\cite{Zhang:2013kja} is achieved by introducing a single auxiliary dimer field that couples to the neutron-core in the $^5P_2$, $^3P_2$ and  $^3P_2^\star$ channels with 3 different couplings. A consequence of this construction is that  
 the amplitudes in the inelastic channels (off-diagonal terms) are specified once the elastic channels (diagonal terms) are known. For example, scattering in the inelastic $^5P_2$-$^3P_2$ channels is specified by the scattering in the elastic  $^5P_2$ and $^3P_2$ channels, respectively. \emph{A priori} there is no known low-energy  symmetry that predicts such a simplification. 
 
In EFT, the kinetic energy terms of the $p$-wave dimer propagators are associated with the corresponding effective momenta. The RG conditions imposed in  deriving Eq.~(15) of Ref.~\cite{Zhang:2013kja}, result in a single effective momentum which is not unexpected since the calculation involved a single dimer field. Therefore, this would predict that the low momentum phase shifts for $^5P_2$, $^3P_2$ and $^3P_2^\star$ channels determined by the same binding momentum $\gamma_0$ and $p$-wave effective momentum $r_1$ are identical~\cite{Zhang:2013kja} up to higher order scattering parameter corrections. Measured $p$-wave phase shifts for $^7\mathrm{Li}(n,n)^7\mathrm{Li}$ and $^7\mathrm{Li}(n,n')^7\mathrm{Li}^\star$ doesn't support this~\cite{Knox:1981dgp}.  In contrast, the coupled-channel calculation presented in subsection~\ref{sec:PwaveMix} involves three separate effective momenta $r_{11}$,  $r_{12}$, $r_{22}$, and the phase shifts in $^3P_2$ and $^3P_2^\star$ channels would differ. The extension to  coupled-channel calculation mixing $^5P_2$-$^3P_2$-$^3P_2^\star$ would involve   6 separate effective momenta, and would result in different phase shifts in $^5P_2$, $^3P_2$ and $^3P_2^\star$ channels, respectively. There is no low-energy symmetry to expect the spin $S=2$ and spin $S=1$ $p$-wave phase shifts to be the same at low momenta up to  corrections that enter the calculation beyond the effective momenta.  

 Two RG conditions were imposed in  Eq.~(15) of Ref.~\cite{Zhang:2013kja} by matching the $T$-matrix for $n$-$^7$Li elastic scattering in the $2^+$ channel to the $p$-wave ERE. The single dimer EFT of Ref.~\cite{Zhang:2013kja} allows for $^5P_2\rightarrow{}^3P_2$ and $^3P_2\rightarrow{}^5P_2$ transitions that were not included. These channels can contribute because the dimer field can connect to these external states though they are not allowed in the self-energy contribution in Eq.~(14) of Ref.~\cite{Zhang:2013kja}. 
 Thus in addition to $h_{(^5P_2)}^2+h_{(^3P_2)}^2$ and $h^2_{(^3P_2^\star)}$, the combination  $h_{(^5P_2)}h_{(^3P_2)}$ should enter the calculation of the $T$-matrix , but not their dimer propagator, in Ref.~\cite{Zhang:2013kja}. In the single dimer formalism, then, the three unknown EFT couplings have to be determined from two RG conditions. We did not explore how this omission affects the renormalization of divergences in the theory and the wave function renormalization constant calculation from the residue of the $T$-matrix. The bound state $T$-matrix calculation for the $1^+$ state in Eq.~(20) of Ref.~\cite{Zhang:2013kja} has similar omissions that affect the RG conditions in Eq. (21). In the coupled-channel calculations we present, all the EFT couplings are determined in terms of scattering parameters through the RG conditions in Eq.~(\ref{eq:pwaveRG}).  There are  other minor technical differences in our calculations such as  we subtract both the linear and cubic divergences for the $p$ waves in PDS~\cite{Rupak:2011nk,Fernando:2011ts} instead of just the linear divergences~\cite{Zhang:2013kja} that appear in the $p$-wave calculations. 

In the incoming $s$-wave channel, the comparison with Ref.~\cite{Zhang:2013kja} is not simple since only the interactions for the elastic  $^3S_1$ (diagonal channel 11) and inelastic ${}^3S_1$-${}^3S_1^\star$ (off-diagonal channel 12 or 21) scatterings are shown. Possible elastic ${}^3S_1^\star$ (diagonal channel 22) scattering, and how the parameters in the three channels due to the mixing relate to each other, are not indicated. 
Zhang \emph{et al.} describe the scattering amplitude in the inelastic $^3S_1$-$^3S_1^\star$ channel by a coupling 
$g_{(^3S_1^\star)}\sim 2\pi/(\mu\Lambda)$. Thus scattering amplitude in the elastic $^3S_1$ and inelastic $^3S_1$-$^3S_1^\star$ channels both scale as  $2\pi/(\mu\Lambda)$ which is the same scaling as ours.
 Moreover, the two calculations find the LO $s$-wave amplitudes to be a constant.
However, this similarity is only superficial. $s$-wave amplitudes at LO with natural-sized  scattering lengths are expected to be constants since they are in $s$ wave. 
We do not know how the single coupling $g_{(^3S_1^\star)}$ and the associated scattering parameters relate to the couplings and parameters in the elastic $^3S_1$ and $^3S_1^\star$ channels. In our expression, we predict $\mathcal A^{(12)}= -a_{22} \mathcal A^{(11)}/a_{12}$ at LO. It is not possible to infer if such a relation also holds in Zhang \emph{et al.}'s calculation. This identification is necessary to interpret the short-distance contribution of the excited $^7\mathrm{Li}^\star$ core explicitly. Moreover, the $Q/\Lambda$ expansion of the amplitudes $\mathcal A^{(11)}$, $\mathcal A^{(12)}$ at small but finite momentum $p$ in Eq.~(\ref{eq:EFTexpansionSwave}) depends on the size of $r_0^{(1)}$ that follows from Eq.~(\ref{eq:EFTstarConstrains}). The scaling of the $s$-wave effective range $r_0^{(1)}$ is not indicated in Ref.~\cite{Zhang:2013kja} which again makes a direct comparison to the present work difficult.

 %==================================================
 \section{A Survey}
 \label{sec:Survey}
 %==================================================
 
 %%------------- Figure-2 --------------------------------
\begin{figure}[tbh]
\begin{center}
\includegraphics[width=0.47\textwidth,clip=true]{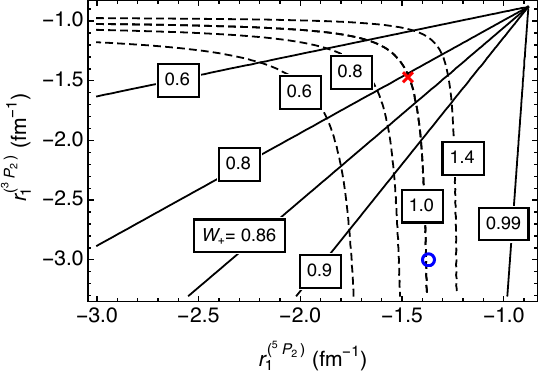}
\end{center}
\caption{\protect 
Correlations in \EFTg. Dashed curves: Contour plot of capture cross section to the $2^+$ state at thermal energy normalized to data~\cite{Lynn:1991}. Solid  lines: Contour plot of branching ratio of capture cross section to the $2^+$ state in the spin $S=2$ channel at thermal energy. The boxed numbers indicate the values on the corresponding contour lines. The $\times$ marks the parameter value $r_1^{(^5P_2)}=\SI{-1.47}{\femto\meter^{-1}}=r_1^{(^3P_2)}$ used in Ref.~\cite{Fernando:2011ts}. The $\circ$ marks $r_1^{(^5P_2)}=\SI{-1.37}{\femto\meter^{-1}}$, 
$r_1^{(^3P_2)}=\SI{-3.0}{\femto\meter^{-1}}$ obtained from fits to ANCs from Ref.~\cite{Trache:2003ir} that are labeled as \EFTg ~ANC in Table~\ref{table:Li8Table}, Fig.~\ref{fig:Survey} and  in the text.} 
\label{fig:Contours}
\end{figure}

 The E1 cross section in Eq.~(\ref{eq:crosssection}) was calculated earlier in Refs.~\cite{Rupak:2011nk, Fernando:2011ts}. It depends on four scattering parameters $a_0^{(2)}$, $a_0^{(1)}$, $r_1^{(^5P_2)}$ and $r_1^{(^3P_2)}$. The two effective momenta are not known experimentally, and the capture to the $2^+$ $^8$Li ground state is sensitive to only a combination of these. 
In the following we define the thermal ratio as the theory calculation of the 
capture cross section to the $2^+$ ${}^8{\rm Li}$ ground state at thermal 
energy, divided by the corresponding experimental value~\cite{Lynn:1991}. 
Fig.~\ref{fig:Contours} shows that a single parameter family of $p$-wave 
effective momenta can reproduce a given thermal ratio (dashed curves). 
The solid  lines show  how a single parameter family of $p$-wave effective momenta can produce a  given branching ratio  $W_+$ of the capture to the $S=2$ channel  at thermal energy. In Refs.~\cite{Rupak:2011nk,Fernando:2011ts} a common effective momentum 
$r_1^{(2^+)}\sim\Lambda$ was used for both  $r_1^{(^5P_2)}$ and $r_1^{(^3P_2)}$ for convenience which reproduced the thermal capture rate. This also gave a branching ratio that is consistent with the experimental lower bound of $W_+\geq 0.86$~\cite{Gulko:1968}, once the theory errors from the \EFTg ~in Ref.~\cite{Fernando:2011ts} are also taken into consideration. A NLO 30\% correction to $r_1^{(^3P_2)}$ is sufficient to satisfy the $W_+$ bound in Fig.~\ref{fig:Contours}. We mention that earlier works have estimated the lower bound as 0.80~\cite{Conner:1959} and 0.75~\cite{Abov:1962}, respectively. Applying the branching ratio lower bound  
$W_+\geq 0.86$ 
to the experimental constraint on thermal capture rate would restrict $r_1^{(^5P_2)}\sim \SI{-1.5}{\femto\meter^{-1}}$ but leave $r_1^{(^3P_2)}\lesssim \SI{-2}{\femto\meter^{-1}}$ 
completely unbound from below. 

%%------------- Figure-3 --------------------------------
\begin{figure}[tbh]
\begin{center}
\includegraphics[width=0.47\textwidth,clip=true]{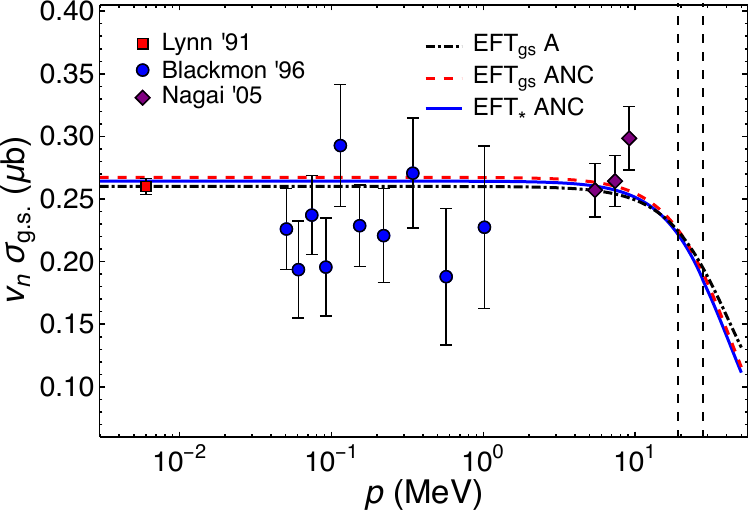}
\includegraphics[width=0.47\textwidth,clip=true]{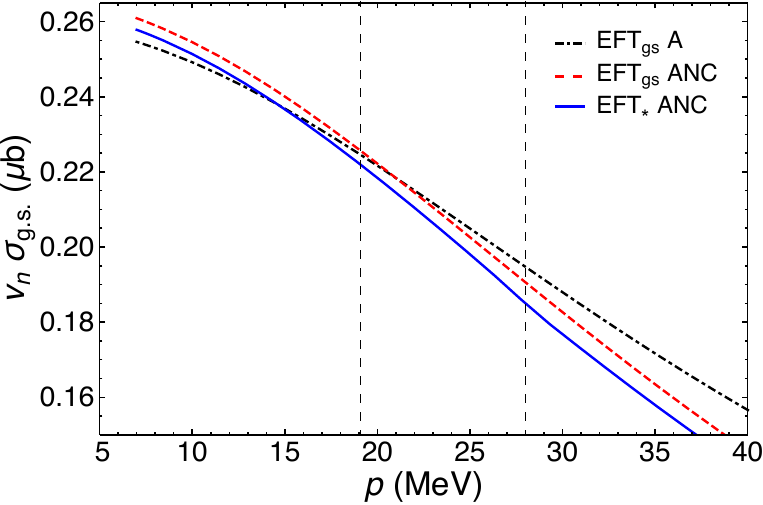}
\end{center}
\caption{\protect Capture to the $2^+$ ground state in the c.m. frame, with $v_n$ the neutron velocity. 
Ref.~\cite{Fernando:2011ts} results without the initial state $d$-wave contribution are shown as the (black) dot-dashed curve for \EFTg ~A, the (red) dashed curve for \EFTg ~ANC, and the solid (blue) curve for EFT${}_\star$ ANC.  The grid lines are at the $3^+$ resonance momentum $p_R=19.1$ MeV and the $^7\mathrm{Li}^\star$ inelasticity $\gamma_\Delta=28.0$ MeV.  }
\label{fig:Survey}
\end{figure}

%------------- Table-1 ---------------------
%\begin{center}
\begin{table*}[tbh]
\centering
\caption{\nLi ~capture to the $2^+$ state. We estimate the parameters as described in the text. Interpretation of $r_1^{(^5P_2)}$ from  $\mathcal Z^{(^5P_2)}$ assumes a single-channel calculation for $S=2$. In \EFTg, a single-channel calculation gives $r_1^{(^3P_2)}=\SI{-1.47+-0.01}{\femto\meter^{-1}}$ and $r_1^{(^3P_2)}=\SI{-3+-0.4}{\femto\meter^{-1}}$ for $\mathcal Z^{(^3P_2)}=\num{7.1+-0.2}$ and $\mathcal Z^{(^3P_2)}=\num{1.9+-0.3}$, respectively.  Thermal ratio is the EFT cross section normalized to Lynn data~\cite{Lynn:1991}. Branching ratio is the capture in the $S=2$ spin channel compared to the total cross section at thermal momentum.
}
%\begin{adjustbox}{width=0.97\textwidth,center}
\begin{ruledtabular}
\begin{tabular}{lccccc}
Theory &$\mathcal Z^{(^5P_2)}$  & $r_1^{(^5P_2)}$ (\si{\femto\meter^{-1}}) & $\mathcal Z^{(^3P_2)}$ 
& Thermal ratio& Branching ratio\\ \hline
\begin{filecontents*}{Li8MixTable.csv}
Theory,ra,dra,Za,dZa,Zb,dZb,ThermalRatio,dTR,BR,dBR
EFT$_\text{gs}$ A,-1.47,0.01,7.1,0.2,7.1,0.2,1,5555,0.809,0.009
EFT$_\text{gs}$ ANC,-1.37,0.04,8.5,0.7,1.9,0.3,1.03,0.08,0.95,0.01
EFT$_\star$ ANC,-1.37,0.04,8.5,0.7,1.9,0.3,1,0.1,0.96,0.05
EFT$_\text{gs}$/EFT$_\star$ Lynn LO,-1.35,0.01,8.8,0.2,9999,5555,1,5555,1,5555
EFT$_\text{gs}$/EFT$_\star$ Lynn-ANC NLO,-1.39,0.01,8.1,0.2,1.9,0.3,1,5555,0.92,0.02
EFT$_\text{gs}$/EFT$_\star$ ANC LO,-1.37,0.04,8.5,0.7,9999,5555,0.98,0.08,1,5555
EFT$_\text{gs}$/EFT$_\star$ ANC NLO,-1.37,0.04,8.5,0.7,1.9,0.3,1.05,0.08,0.92,0.02
\end{filecontents*}
\csvreader[head to column names, late after line=\\]{Li8MixTable.csv}{}
{\ \Theory 
&
\ifthenelse{\equal{\Za}{9999}}{---}{
\ifthenelse{\equal{\dZa}{5555}}{$\Za$}
{\SI{\Za+-\dZa}{}}
}
& 
\ifthenelse{\equal{\ra}{9999}}{---}{
\ifthenelse{\equal{\dra}{5555}}{$\ra$}
{\SI{\ra+-\dra}{}}
}
&
\ifthenelse{\equal{\Zb}{9999}}{---}{
\ifthenelse{\equal{\dZb}{5555}}{$\Zb$}
{\SI{\Zb+-\dZb}{}}
}
&\ifthenelse{\equal{\dTR}{5555}}{\ThermalRatio}
{\SI{\ThermalRatio+-\dTR}{}}
&\ifthenelse{\equal{\dBR}{5555}}{\BR}
{\SI{\BR+-\dBR}{}}
}
\end{tabular}
\end{ruledtabular}
%\end{adjustbox}
 \label{table:Li8Table}
\end{table*}

The result from Ref.~\cite{Fernando:2011ts}, without initial state $d$-wave contribution, is plotted as \EFTg ~A in Fig.~\ref{fig:Survey}, where data from Refs.\cite{Lynn:1991,Blackmon:1996,Nagai:2005} are also shown. As pointed out in Ref.~\cite{Zhang:2013kja}, one could use the known ANCs $C_{1,\zeta}^2$ to fit the effective momenta using the relation:
 %--------------- Equation-32 -----------------
 \begin{align}
     C_{1,\zeta}^2=\frac{\gamma^2}{\pi} \frac{2\pi}{\mu} \mathcal Z^{(\zeta)}\, ,
 \end{align}
 where $\gamma$ is the appropriate binding momentum in the channel $\zeta$.  We use the following ANC values: $C_{1,^5P_2}^2=\SI{0.352+-0.028}{\femto\meter^{-1}}$, $C_{1,^3P_2}^2=\SI{0.08+-0.013}{\femto\meter^{-1}}$, $C_{1,^5P_1}^2=\SI{0.047+-0.004}{\femto\meter^{-1}}$, 
 $C_{1,^3P_1}^2=\SI{0.035+-0.005}{\femto\meter^{-1}}$ 
 from neutron transfer reaction~\cite{Trache:2003ir} and 
 $C_{1,^3P_2^\star}^2=\SI{0.147+-0.005}{\femto\meter^{-1}}$, 
 $C_{1,^3P_1^\star}^2=\SI{0.0458+-0.001}{\femto\meter^{-1}}$  from an \emph{ab initio} calculation~\cite{Zhang:2013kja}.
$C_{1,^5P_2}^2$, $C_{1,^3P_2}^2$, $C_{1,^5P_1}^2$, $C_{1,^3P_1}^2$  were calculated from the measured~\cite{Trache:2003ir} $C_{1,p_{3/2}}^2=\SI{0.384+-0.038}{\femto\m^{-1}}$,  $C_{1,p_{1/2}}^2=\SI{0.048+-0.006}{\femto\m^{-1}}$,  $C_{1,p_{3/2}^\star}^2=\SI{0.067+-0.007}{\femto\m^{-1}}$,  $C_{1,p_{1/2}^\star}^2=\SI{0.015+-0.002}{\femto\m^{-1}}$ using Eqs.~(\ref{eq:2plusstate}), ~(\ref{eq:1plusstate}): 
$C_{1,^3P_2}= \frac{1}{\sqrt{2}}[C_{1,p_{3/2}}-C_{1,p_{1/2}}]$, $C_{1,^5P_2}=\frac{1}{\sqrt{2}} [C_{1,p_{3/2}}+C_{1,p_{1/2}}]$, $C_{1,^3P_1}= \frac{1}{\sqrt{6}}[\sqrt{5}C_{1,p_{3/2}^\star}-C_{1,p_{1/2}^\star}]$, 
$C_{1,^5P_1}= \frac{1}{\sqrt{6}}[C_{1,p_{3/2}^\star}+\sqrt{5}C_{1,p_{1/2}^\star}]$. The signs of $C_{1,p_{3/2}}$, $C_{1,p_{1/2}}$, $C_{1,p_{3/2}^\star}$, $C_{1,p_{1/2}^\star}$ were picked to be compatible with \emph{ab initio}  calculations~\cite{Zhang:2013kja,Nollett:2011qf}.

 The experimental ANCs from Ref.~\cite{Trache:2003ir} give  $r_1^{(^5P_2)}\sim\SI{-1.37}{\femto\meter^{-1}}$, $r_1^{(^3P_2)}\sim\SI{-3.0}{\femto\meter^{-1}}$, shown in Fig.~\ref{fig:Contours}, which are consistent with the expectation from the analysis above. The corresponding result for the capture cross section is plotted as \EFTg ~ANC in Fig.~\ref{fig:Survey}. Though the curves \EFTg ~A and \EFTg ~ANC differ, this difference is within the expected theoretical error. We provide a more robust theory error estimate later in Section~\ref{sec:powercounting} when we develop the EFT power counting relevant to E1 transition at low momenta. Table~\ref{table:Li8Table} lists some of the fit parameters, and the thermal and branching ratios. 
 
 In Fig.~\ref{fig:Survey} we also plot the capture calculation to the $2^+$ state in EFT$_\star$. It is labeled as EFT$_\star$ ANC. 
In the spin $S=2$ channel this cross section depends on $a_0^{(2)}$ and  $\mathcal Z^{(^5P_2)}$.
In the $S=1$ channel 
we need the parameters $a_0^{(1)}$, $a_{12}$, $a_{22}$, and overall normalization constants $\mathcal Z^{(^3P_2)}$, $\mathcal Z^{(^3P_2^\star)}$. The unknowns $\mathcal Z^{(^5P_2)}$, $\mathcal Z^{(^3P_2)}$ are constrained from the measured ANCs~\cite{Trache:2003ir}, and $\mathcal Z^{(^3P_2^\star)}$ is constrained from the calculated ANC~\cite{Zhang:2013kja}. 
We take $a_{22}=-1/200(1\pm0.4)\,\si{MeV^{-1}}$ and $a_{12}=1/200(1\pm0.4)\,\si{MeV^{-1}}$. The result shows a correlation between $a_{22}$ and $a_{12}$, and other sign choices for these two scattering lengths vary the result a few percent (well within a NNLO contribution we discuss later).  We picked the combination that gives a better description of the thermal data but it is not crucial for the analysis. In this plot we did not approximate the $s$-wave scattering amplitudes $\mathcal A^{(11)}$, $\mathcal A^{(12)}$ in Eq.~(\ref{eq:EFTexpansionSwave}) by the $Q/\Lambda$ expansion. We use the central values of the parameters for the plots. A more careful accounting of the input error is given later.
One notices that the curves \EFTg ~ANC and EFT${}_\star$ ANC are compatible with each other within the expected errors regarding their momentum dependence. 
In particular, the opening of the inelastic threshold at $p=\gamma_{\Delta}=\SI{28.0}{\mega\eV}$ remains a small effect. 
Nevertheless, the range of momentum where the explicit ${}^7{\rm Li}_\star$ degree of freedom gives the largest contributions is where the M1 capture due to the $3^{+}$ resonance is already predominant~\cite{Fernando:2011ts}. Given that $p_R<\gamma_\Delta$, the inclusion of the $3^{+}$ degree of freedom precedes that of the ${}^7{\rm Li}_\star$ in the formulation of the low-energy effective theory for the ${}
^7{\rm Li}(n,\gamma)^8{\rm Li}$ reaction if we attempt to describe data. Still, one can analyze the non-resonant capture theoretically. 
We interpret the plot of EFT$_\star$ ANC as indicative of the fact that the $^7\mathrm{Li}^\star$ contribution to the initial $s$-wave scattering  at these energies is a subleading effect. The contribution of the $^7\mathrm{Li}^\star$ to the bound state wave function cannot be separated completely  with the two ANCs for $^3P_2$ and $^3P_2^\star$ states to constrain the three relevant $p$-wave effective momenta $r_{11}$, $r_{12}$, and $r_{22}$ in Eq.~(\ref{eq:ZphiMix}).
The fitted $\mathcal Z^{(^5P_2)}$ can be interpreted in terms of  $r_1^{(^5P_2)}$ in  Table~\ref{table:Li8Table}, if treated as a single-channel calculation. 
In Table~\ref{table:Li8Table}, we also list the thermal and branching ratios for the EFT$_\star$ ANC fit.

 %==================================================
 \section{\texorpdfstring{\EFTgsHead}{EFTgs} and \texorpdfstring{\EFTstarHead}{EFT*} Power Counting}
 \label{sec:powercounting}
 %==================================================
 
 We start with the capture to the dominant $2^+$ state. Capture in the spin $S=2$ channel is about 4 times larger than in the spin $S=1$ channel~\cite{Barker:1995}. If we consider the branching ratio for capture in the spin $S=2$ channel to the $2^+$ ground  
 state in \EFTg ~at threshold, then we can analytically calculate~\cite{Rupak:2011nk} from Eq.~(\ref{eq:crosssection})
 %--------------- Equation-33 -----------------
\begin{multline}
    \frac{\sigma^{(^5P_2)}}{\sigma^{(^5P_2)}+\sigma^{(^3P_2)}}=\\\frac{(3-2 a_0^{(2)}\gamma_0)^2 \mathcal Z^{(^5P_2)}}{(3-2 a_0^{(2)}\gamma_0)^2 \mathcal Z^{(^5P_2)}+(3-2 a_0^{(1)}\gamma_0)^2 \mathcal Z^{(^3P_2)}}\, ,
\end{multline}
which indicates that the dominance of the $S=2$ channel capture is due to a combination of the larger numerical value of $|a_0^{(2)}|\gg a_0^{(1)}$, and also the respective signs of the scattering lengths, even if we take  $\mathcal Z^{(^5P_2)}\sim  \mathcal Z^{(^3P_2)}$. The calculated $\mathcal Z^{(^5P_2)}\gg \mathcal Z^{(^3P_2)}$ in Table~\ref{table:Li8Table} makes the branching ratio even larger. In EFT, we can only assume numerical sizes for the $s$-wave scattering lengths and not make prediction about their overall signs. Thus to develop a power counting, we have to explicitly account for the empirical fact that capture in $S=2$ dominates which requires more than the scaling $|a_0^{(2)}|\sim 1/Q\gg a_0^{(1)}\sim 1/\Lambda$.  
We will take the dominance of capture in spin $S=2$ channel as given in addition to the scaling of the $s$-wave scattering length.
 Thus we count capture in the $S=1$ channel to be NLO (a subleading effect). In the dominant spin channel, the $s$-wave scattering length $|a_0^{(2)}|\sim 1/Q$ is large. At low momentum the contribution from the initial state interaction scales as $a_0^{(2)} (B_0+J_0)$. The loop integral combination scales as $B_0+J_0\sim Q$.  Thus the contributions from the diagrams with and those without initial state strong interactions in Fig.~\ref{fig:Capture} are of the same size $\mathcal O(1)$ in the $Q/\Lambda$ expansion. This constitutes the LO contribution in both \EFTg ~and EFT$_\star$.
 
 The cross section in the $S=2$ channel is proportional to the wave function renormalization constant $\mathcal Z^{(^5P_2)}$ that depends on the binding momentum $\gamma_0\sim Q$ and effective momentum $r_1^{(^5P_2)}\sim \Lambda$. Ideally, one expands $\mathcal Z^{(^5P_2)}$ in $\gamma_0/r_1^{(^5P_2)}\sim Q/\Lambda$. However, the actual expression in Eq.~(\ref{eq:Zphi}) would involve an expansion in $3\times\gamma_0/|r_1^{(^5P_2)}|\lesssim 1$ that converges slowly, see Table~\ref{table:Li8Table}. We just resum the entire series since the exact expression is known~\cite{Rupak:2011nk,Fernando:2011ts}. In Eq.~(\ref{eq:Zphi}), there are no corrections from higher order $p$-wave scattering parameters once the ERE parameters are expanded around the pole at $p=i\gamma_0$ instead of a Taylor series around $p=0$, similar to Eq.~(\ref{eq:poleExp}) but for a single-channel. Alternatively, a compromise would be to adhere to a strict $Q/\Lambda$ expansion but use the so called  zed-parametrization. It was introduced in Ref.~\cite{Phillips:1999hh} for $s$-wave bound states. Ref.~\cite{Fernando:2011ts} extended the formalism to $p$-wave bound states though didn't apply it to the actual calculation, choosing to instead resum the $3\gamma_0/r_1^{(^5P_2)}$ series. The zed-parametrization in $^5P_2$ channel would recover the exact result at NLO at the cost of introducing large corrections to the LO result. 
 
At NLO, there is an $s$-wave effective range $r_0^{(2)}\sim 1/\Lambda$ correction. The capture data at low energy is not sensitive to this parameter. Fig.~\ref{fig:2pluscapture} shows the sensitivity to $r_0^{(2)}$ in the range \SIrange{-5}{5}{\femto\meter}. We include this in our error estimates later in the analysis. 
 
 The capture in the spin $S=1$ channel starts at NLO, and the power counting for \EFTg ~and EFT$_\star$  has to be discussed separately though  they have the same momentum dependence up to NLO. We start with \EFTg ~where  the contributions from Figs.~\ref{fig:Capture} $(b)$, $(d)$ 
from the initial state interactions scale as $a_0^{(1)} (B_0+J_0)$. 
Given the smaller $a_0^{(1)}\sim 1/\Lambda$ in this channel, these initial 
state interactions are $Q/\Lambda$ suppressed compared to the contributions 
from Figs.~\ref{fig:Capture} $(a)$, $(c)$, thus they constitute a NNLO 
contribution. 
The wave function renormalization constant $\mathcal Z^{(^3P_2)}$ has the same form as Eq.~(\ref{eq:Zphi}) for the $S=2$ channel.
 
 In EFT$_\star$ there are two differences in the $S=1$ channel from the previous discussion due to the mixing in the initial $^3S_1$-$^3S^\star_1$ scattering state and final $^3P_2$-$^3P^\star_2$ bound state. The scattering state contribution scales as 
$\mathcal A^{(11)} (B_0+J_0)$, $\mathcal A^{(12)} (B_0+J_0)$. 
In the $S=1$ channel, $a_0^{(1)}\sim 1/\Lambda$ and assuming all the scattering length parameters $a_{11}\sim a_{12}\sim a_{22}\sim 1/\Lambda$ to be natural as well, we see from Eq.~(\ref{eq:EFTexpansionSwave}) that the initial state interaction in this channel also scales as $Q/\Lambda$ compared to the contributions Fig.~\ref{fig:Capture} $(a)$, $(c)$ 
without initial state interaction. Thus in this theory also the scaling for the $s$-wave interaction is similar to \EFTg ~up to NLO. There is one difference, however, from before. Now    we have two wave function renormalization constants $\mathcal Z^{(^3P_2)}$, $\mathcal Z^{(^3P_2^\star)}$ that are shown in Eq.~(\ref{eq:ZphiMix}). Given that the ANCs for $^3P_2$ and $^3P_2^\star$~\cite{Zhang:2013kja, Nollett:2011qf} are of similar size, we do not attempt any perturbative expansion in the mixing parameter $r_{12}$. 
 
 To summarize, the LO contribution to the capture to the $2^+$ ground state is from the spin $S=2$ channel. At this order the cross section depends only on the $^5S_2$ scattering length $a_0^{(2)}$ and  the wave function renormalization constant $\mathcal Z^{(^5P_2)}$ in both  \EFTg ~and EFT$_\star$. 
The NLO corrections come from the $^5S_2$ effective range $r_0^{(2)}$ and from the capture in the spin $S=1$ channel without initial interaction in either of the two theories, \EFTg ~or EFT$_\star$. 
Thus the momentum dependence in \EFTg ~and EFT$_\star$ are indistinguishable at NLO. Up to this order, the difference between the two theories lies in the interpretation of the wave function renormalization constant $\mathcal Z^{(^3P_2)}$---in \EFTg ~one directly relates $\mathcal Z^{(^3P_2)}$ with the $p$-wave effective momentum $r_1^{(^3P_2)}$. In EFT$_\star$, $\mathcal Z^{(^3P_2)}$  (and $\mathcal Z^{(^3P_2^\star)}$) is a function of three effective momenta $r_{11}$, $r_{12}$, $r_{22}$. The NLO result depends only on $\mathcal Z^{(^5P_2)}$, $\mathcal Z^{(^3P_2)}$, and once they are fitted to capture data and/or ANCs, \EFTg ~and  EFT$_\star$ results are indistinguishable.

The proposed power counting for the capture to the $1^+$ state of $^8$Li in this work is very similar to the one proposed above for the $2^+$ ground state. The excited $^8$Li binding momentum $\gamma_1$ is relatively smaller and effective momentum $r_1^{(^5P_1)}$ is larger than $\gamma_0$ and $r_1^{(^5P_2)}$ of the $^8$Li ground state, respectively, see Table~\ref{table:Li8TableB}.  In $^5P_1$ channel we could use the zed-parametrization with a smaller NLO contribution. To keep the presentation simple, we resum and use the exact $\mathcal Z^{(^5P_1)}$ at LO, similar to the $\mathcal Z^{(^5P_2)}$ calculation.

We end this section with a discussion of higher order corrections. The three main contributions to the capture cross sections, discussed earlier in section~\ref{sec:interaction}, are from initial state scattering, electromagnetic currents, and final bound state calculation. The contribution from the bound state calculation is contained in the wave function renormalization constants $\mathcal Z$s whose expressions are known exactly in terms of the $p$-wave effective momenta. These do not receive corrections from higher order scattering parameters. We simply fit the $\mathcal Z$s to capture data and/or ANCs, and consequently no theory error is associated with the bound state calculation. 

At the momenta $p\lesssim\gamma_0$ that we consider, E1 transition is the most relevant one. M1 transition was found to be relevant only around the $3^+$ resonant momentum $p_R=19.1$ MeV~\cite{Fernando:2011ts}. The E2 transition strength was estimated in Ref.~\cite{Izsak:2013hga} to be several orders suppressed compared to the E1 strength. The E2 capture cross section is suppressed by additional factor of $k_0^4$ compared to the E1 capture when relating cross sections to electromagnetic transition strengths~\cite{Bertulani,RADCAP},  and thus neglected here.

 For capture to the $2^+$ ground state one can 
 consider  in \EFTg ~generic two-body E1 operators $(e Z_c/M_c) L_2 [\chi_{ij}^{(^5P_2)}]^\dagger E_k \chi_{xy}^{(^5S_2)} T_{ijkxy}$ and $(e Z_c/M_c) L_1 [\chi_{ij}^{(^3P_2)}]^\dagger E_x \chi_{y}^{(^3S_1)} R_{ijxy}$ in spin $S=2$ and $S=1$ channels, respectively. 
  $\bm{E}$ is the electric field, and $T_{ijkxy}$, $R_{ijxy}$ are defined in  Appendix~\ref{sec:Projectors}. The dimensionless couplings $L_i$ are defined to be compatible with the dimer-neutron-core coupling in Eq.~(\ref{eq:EFT}). For example, the capture from $^5S_2$ involves large rescattering due to the large numerical value of the $s$-wave scattering length, and a direct calculation gives an amplitude proportional to  
 $-(e Z_c/M_c) \mathcal D(E,0) k_0 L_2 \sqrt{2\pi\mathcal Z^{(^5P_2)}/\mu}$ with $k_0=(p^2+\gamma_0^2)/(2\mu)$ the photon energy. From Eq.~(\ref{eq:swaveAmp}) one notices that, for ${p\lesssim 1/|a_0^{(2)}|}$, one has ${\mathcal D(E,0) \sim a_0^{(2)}}$ and the two-body current gives a relative contribution $a_0 k_0 L_2$ [factoring out the common $(e Z_c/M_c)\sqrt{2\pi\mathcal Z^{(^5P_2)}/\mu}$] to the  amplitude. In Refs.~\cite{Rupak:2011nk,Fernando:2011ts}, $k_0$ was estimated to scale as $Q^2/\Lambda$, however, given the large value $\mu\gg\Lambda$ it is numerically more reasonable to assume $\Lambda/\mu\sim Q/\Lambda$~\cite{Higa:2008dn,Higa:2016igc}, thus $k_0\sim Q^3/\Lambda^2$. Using $|a_0^{(2)}|\sim 1/Q$ we estimate the relative two-body current contribution $a_0 k_0 L_2\sim Q^2/\Lambda^2$, a NNLO effect for a natural-sized ${L_2\sim 1}$.   This estimate of the two-body current is one order higher in perturbation than the estimate in Refs.~\cite{Rupak:2011nk,Fernando:2011ts} due to the different scaling of $k_0$. 
 
 One can integrate out the auxiliary $^5S_2$ dimer field without any physical consequences. From Eq.~(\ref{eq:EFT}), replacing the dimer field  through the equation of motion, $(e Z_c/M_c) L_2 [\chi_{ij}^{(^5P_2)}]^\dagger E_k \chi_{xy}^{(^5S_2)} T_{ijkxy}\rightarrow -(e Z_c/M_c)\sqrt{2\pi/\mu} L_2 [\chi_{ij}^{(^5P_2)}]^\dagger E_k (N Q_{ij}C) T_{ijkxy}/\Delta^{(^5S_2)}$ where $Q_{ij}$ is the appropriate Clebsch-Gordan coefficient matrix defined in Appendix~\ref{sec:Projectors}. The contribution from this two-body operator written without the dimer field is straightforward to calculate. One gets $-(e Z_c/M_c) (k_0 L_2/ \Delta^{(^5S_2)}) \sqrt{2\pi\mathcal Z^{(^5P_2)}/\mu}/[1-2\pi J_0(-ip)/(\mu\Delta^{(^5S_2)})]$ where the loop integral $J_0(-ip)$ was defined earlier in Eq.~(\ref{eq:swaveSigma}). Application of the RG conditions similar to Eq.~(\ref{eq:swaveRG}) for single-channel results in the same relative contribution  $a_0 k_0 L_2$ at low momentum in the theory without $^5S_2$ dimer as expected.
 
 Two-body current contributions from initial $^3S_1$ channel are even smaller. The scattering length $a_0^{(1)}\sim 1/\Lambda$ in this channel does not give the enhancement we get from $a_0^{(2)}\sim 1/Q$. Unlike the $^5S_2$ channel, the strong interaction is perturbative and one expands in $p a_0^{(1)}\sim Q/\Lambda$. Moreover, the capture in spin $S=1$ is one order suppressed so two-body currents in this channel are beyond NNLO. In EFT$_\star$, mixing in the final $p$-wave bound state doesn't change the power counting estimate of the two-body currents as it only affects the calculation of the $\mathcal Z$s. Moreover, mixing in  $^3S_1$-$^3S_1^\star$ does not result in a large scattering length in our power counting. The $S=1$ channel remains subleading and the corresponding two-body currents in EFT$_\star$ are beyond NNLO.

 E1 capture can proceed from both initial $s$ and $d$ waves.   The $d$-wave contribution without strong interaction, relative to the $s$-wave contribution, is found~\cite{Rupak:2011nk,Fernando:2011ts} to scale as 
 $p^2/(p^2+\gamma^2)$ in the capture amplitude which, though formally $\mathcal O(1)$ for $p\sim\gamma$,  is kinematically suppressed at low momentum. 
One takes $\gamma=\gamma_0$, $\gamma_1$ as appropriate for capture to the $2^+$ and $1^+$ $^8$Li states, respectively. At $p\gtrsim 40$ MeV, this contribution is around 10\% for capture to the $2^+$ ground state in the dominant $S=2$ channel, and we count it as a NNLO contribution. Strong interaction in initial $d$ waves should be kinematically suppressed further as such interaction vertices for elastic scattering necessarily involve at least four (two for each incoming and outgoing channels) extra factors of momentum $p$.

The initial $s$-wave state comprises the total spin channels $S=2$ and $S=1$. The LO and NLO contributions in the $^5S_2$ channel, that has a large scattering length $a_0^{(2)}\sim 1/Q$, are already included. 
The NNLO correction from the ERE is proportional to $[a_0^{(2)} r_0^{(2)} p^2/2]^2\sim (Q/\Lambda)^2$~\cite{Chen:1999tn} and, like the $d$-wave contribution, is kinematically suppressed at low momentum.  The shape parameter $\mathcal P_0^{(2)}$ enters at N$^3$LO.  The  strong interaction in the $^3S_1$ channel in both \EFTg ~and EFT$_\star$, with the smaller scattering length $a_0^{(1)}\sim 1/\Lambda\ll |a_0^{(2)}|$, enters at NNLO in our power counting.

 %==================================================
 \section{Results and Analysis}
 \label{sec:Results}
 
 %%------------- Figure-4 --------------------------------
\begin{figure}[tbh]
\begin{center}
\includegraphics[width=0.47\textwidth,clip=true]{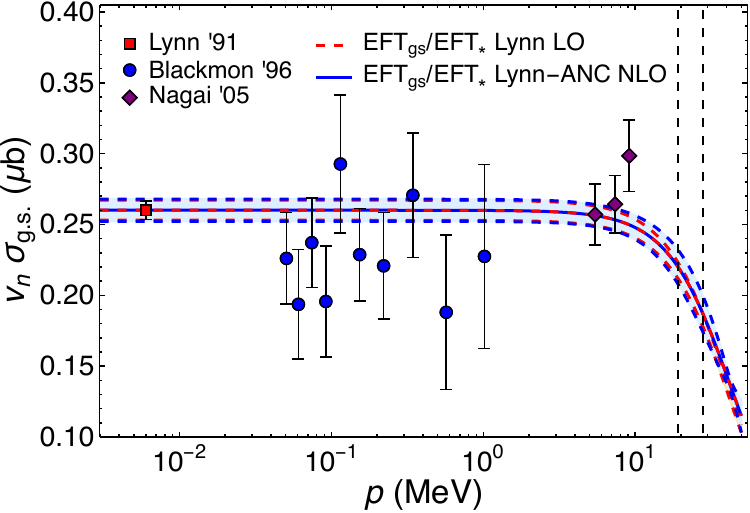}
\includegraphics[width=0.47\textwidth,clip=true]{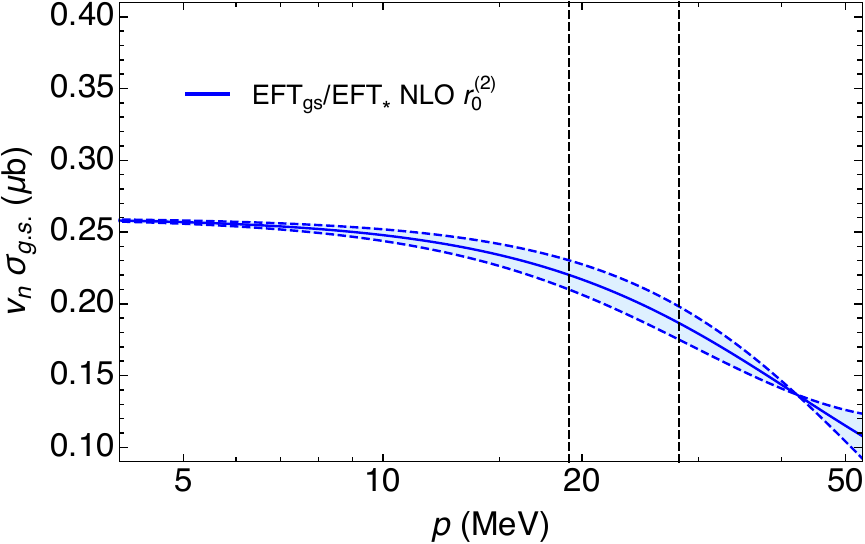}
\end{center}
\caption{\protect Capture to the $2^+$ state in the c.m. frame. Up to  NLO, \EFTg ~and EFT$_\star$ are equivalent. Top panel: The LO (red) dashed  and NLO (blue) solid curves, and error bands (red dashed and blue dashed curves) from inputs overlap. They were constrained by thermal capture data~\cite{Lynn:1991}.
We varied $r_0^{(2)}$ in the range \SIrange{-5}{5}{\femto\meter}. The bottom panel shows only the variation due to $r_0^{(2)}$. The grid lines are explained in Fig.~\ref{fig:Survey}.}
\label{fig:2pluscapture}
\end{figure}
 
In Fig.~\ref{fig:2pluscapture}, the 
\EFTg/EFT$_\star$ Lynn LO 
curve was generated by constraining the wave function renormalization constant $\mathcal Z^{(^5P_2)}$ from the thermal capture data~\cite{Lynn:1991}.  The corresponding effective momentum $r_1^{(^5P_2)}$ value from Eq.~(\ref{eq:Zphi}) is in Table~\ref{table:Li8Table}. 
We took $a_0^{(2)}=\SI{-3.63+-0.05}{\femto\meter}$~\cite{Koester:1983}. 
At NLO, the wave function renormalization $\mathcal Z^{(^3P_2)}$ was constrained by the ratio of the ANCs $C_{1,^3P_2}^2/C_{1,^5P_2}^2=\num{0.228+-0.042}$~\cite{Trache:2003ir}, and then $\mathcal Z^{(^5P_2)}$ was fitted to thermal capture data.
This is the curve labeled \EFTg/EFT$_\star$ Lynn-ANC NLO 
in Fig.~\ref{fig:2pluscapture}. 
As discussed earlier, see Fig.~\ref{fig:Contours}, it is important to independently constrain the wave function renormalization constants in the $S=2$ and $S=1$ channels since the capture data and the branching ratio are not sufficient.   $\mathcal Z^{(^3P_2)}$ can be expressed in terms of a single effective momentum in \EFTg ~but in EFT$_\star$ it depends on three effective momenta. 
As mentioned earlier, 
we do not attempt to write $\mathcal Z^{(^3P_2)}$ in terms of these effective momenta. The LO and NLO curves and the error bands, from errors in the input parameters only,  overlap. This is primarily a consequence of constraining both the LO and NLO results to the same thermal capture data. The errors in Fig.~\ref{fig:2pluscapture} were propagated in quadrature from the errors in the input parameters. 

In the bottom panel of Fig.~\ref{fig:2pluscapture}, we show only the error associated with varying the $^5S_2$ channel effective range $r_0^{(2)}\sim 1/\Lambda$ between \SIrange{-5}{5}{\femto\meter}. $r_0^{(2)}$ has a noticeable impact on the capture cross section only at larger momenta where precision data is lacking, and also happens to be in a momentum range where the $3^+$ resonance contribution 
is significant. 
The resonance contribution can be added in EFT as discussed earlier~\cite{Fernando:2011ts}. In this work we only include the non-resonant contribution. 
 
 %%------------- Figure-5 --------------------------------
\begin{figure}[tbh]
\begin{center}
\includegraphics[width=0.47\textwidth,clip=true]{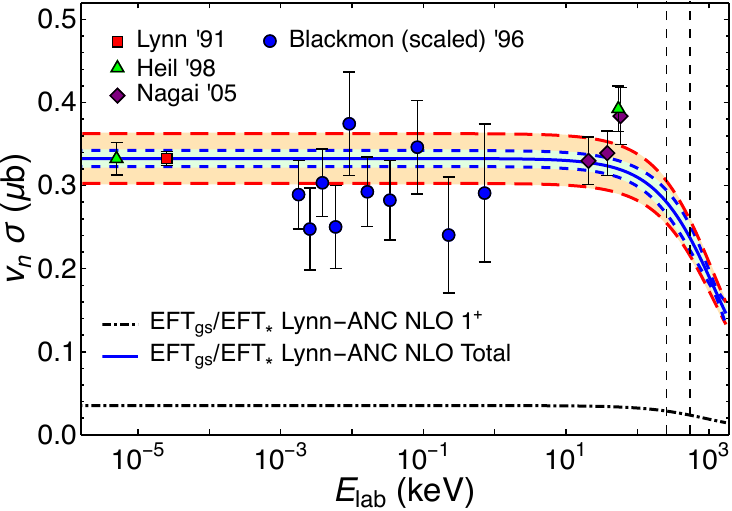}
\end{center}
\caption{\protect Total capture to the $2^+$ and $1^+$ state at NLO  constrained by the thermal capture data~\cite{Lynn:1991} in the lab frame.
We varied $r_0^{(2)}$ in the range \SIrange{-5}{5}{\femto\meter}. The (blue) shaded region between the (blue) dashed curves shows the error due to the input parameters. The (reddish) band between the (red) long-dashed  curves indicate the 10\% NNLO EFT errors. The (blue) solid curve is the total NLO result and the (black) dot-dashed curve is the NLO capture to the $1^+$ state. 
The error bands in the capture to the $1^+$ state are not shown but included in the total capture cross section. The grid lines are explained in Fig.~\ref{fig:Survey}. Ref.~\cite{Blackmon:1996} measured capture to the $2^+$ ground state which was scaled in that work by a factor of 1/0.894 to represent the total capture rate. }
\label{fig:TotalCapture}
\end{figure}

%------------- Table-2 ---------------------
%\begin{center}
\begin{table*}[tbh]
\centering
\caption{\nLi ~capture to the $1^+$ state. We estimate the parameters as described in the text. 
Interpretation of $r_1^{(^5P_1)}$ from  $\mathcal Z^{(^5P_1)}$, assumes a single-channel calculation for $S=2$. 
Thermal ratio is the EFT cross section normalized to Lynn data~\cite{Lynn:1991}. Branching ratio is capture in the $S=2$ spin channel compared to the total cross section at thermal momentum. We corrected the $p$-wave  effective momenta values from Ref.~\cite{Fernando:2011ts} in \EFTg ~A below.}
%\begin{adjustbox}{width=0.97\textwidth,center}
\begin{ruledtabular}
\begin{tabular}{lccccc}
Theory & $\mathcal Z_1^{(^5P_1)}$  & $r_1^{(^5P_1)}$ (\si{\femto\meter^{-1}}) 
& $\mathcal Z_1^{(^3P_1)}$ 
& Thermal ratio& Branching ratio\\ \hline
\begin{filecontents*}{Li8MixTableExc.csv}
Theory,ra,dra,Za,dZa,Zb,dZb,ThermalRatio,dTR,BR,dBR
EFT$_\text{gs}$ A,-2.4,0.4,2.4,0.5,2.4,0.5,1,5555,0.94,0.02
EFT$_\text{gs}$/EFT$_\star$ Lynn LO,-2.3,0.2,2.6,0.3,9999,5555,1,5555,1,5555
EFT$_\text{gs}$/EFT$_\star$ Lynn-ANC NLO,-2.4,0.2,2.4,0.3,1.8,0.3,1,5555,0.94,0.02
EFT$_\text{gs}$/EFT$_\star$ ANC LO,-2.5,0.2,2.2,0.2,9999,5555,0.87,0.07,1,5555
EFT$_\text{gs}$/EFT$_\star$ ANC NLO,-2.5,0.2,2.2,0.2,1.6,0.2,0.92,0.07,0.94,0.01
\end{filecontents*}
\csvreader[head to column names,late after line=\\]{Li8MixTableExc.csv}{}
{\ \Theory 
&
\ifthenelse{\equal{\Za}{9999}}{---}{
\ifthenelse{\equal{\dZa}{5555}}{$\Za$}
{\SI{\Za+-\dZa}{}}
}
& 
\ifthenelse{\equal{\ra}{9999}}{---}{
\ifthenelse{\equal{\dra}{5555}}{$\ra$}
{\SI{\ra+-\dra}{}}
}
&
\ifthenelse{\equal{\Zb}{9999}}{---}{
\ifthenelse{\equal{\dZb}{5555}}{$\Zb$}
{\SI{\Zb+-\dZb}{}}
}
&\ifthenelse{\equal{\dTR}{5555}}{\ThermalRatio}
{\SI{\ThermalRatio+-\dTR}{}}
&\ifthenelse{\equal{\dBR}{5555}}{\BR}
{\SI{\BR+-\dBR}{}}
}
\end{tabular}
\end{ruledtabular}
%\end{adjustbox}
 \label{table:Li8TableB}
\end{table*}  

The plot in Fig.~\ref{fig:TotalCapture} includes capture to both the $2^+$ ground   and $1^+$ excited states of $^8$Li at NLO. The $2^+$ capture cross section is the same as in Fig.~\ref{fig:2pluscapture}.  The wave function renormalization constant $\mathcal Z^{(^3P_1)}$ in the $1^+$ capture is constrained from the ratio $C_{1,^3P_1}^2/C_{1,^5P_1}^2=\num{0.73+-0.12}$~\cite{Trache:2003ir}, and $\mathcal Z^{(^5P_1)}$ from the thermal capture data~\cite{Lynn:1991}. We include the expected 10\% \EFTg/EFT$_\star$ error in Fig.~\ref{fig:TotalCapture} from NNLO corrections. 
We show some parameters for the capture to $1^+$ in Table~\ref{table:Li8TableB}. As in the $2^+$ capture,
the $S=2$ channel effective momentum $r_1^{(^5P_1)}$ is calculated 
from the wave function renormalization constant $\mathcal Z^{(^5P_1)}$ using a relation similar to Eq.~(\ref{eq:Zphi}), treating it as a single-channel calculation. We see that the extrapolation of the \EFTg/EFT$_\star$ curves to low momentum gives an accurate postdiction of the sub-thermal datum~\cite{Heil:1998}. 
A few numerical results are shown in Table~\ref{table:Li8TlabTable}.

The higher momentum data points in Fig.~\ref{fig:TotalCapture} from Refs.~\cite{Nagai:2005, Heil:1998} are associated with the M1 transition from the initial $3^+$
resonance state. As mentioned before, this contribution can be included in EFT as presented in Ref.~\cite{Fernando:2011ts} where the resonance is described as a $^5P_3$ state of the valence neutron and the ground state of $^7$Li. The excited state $^7\mathrm{Li}^\star$ does not contribute to the initial ${}^5P_3$ scattering state. To keep the discussion more focused we do not include the M1 contribution.

%------------- Table-3 ---------------------
%\begin{center}
\begin{table}[tbh]
\centering
\caption{\nLi ~E1 capture  to the $2^+$ and $1^+$ state. These correspond to the NLO \EFTg/EFT$_\star$ result from Fig.~\ref{fig:TotalCapture}. The errors from the inputs are 
propagated in quadrature. The estimated 10\% NNLO theory error is not shown.
}
%\begin{adjustbox}{width=0.97\textwidth,center}
\begin{ruledtabular}
\begin{tabular}{rll}
$E_\text{lab}$ (\si{\kilo\eV}) &$v_n \sigma_\text{E1}^{(2^+)}$ (\si{\mu\barn})  & $v_n \sigma_\text{E1}^{(1^+)}$ (\si{\mu\barn})\\ \hline
\begin{filecontents*}{Li8TlabTable.csv}
Elab,sigmaA,dsigmaAA,dsigmaAB,sigmaB,dsigmaBA,dsigmaBB
0,0.297,0.009,0.03,0.0352,0.0037,0.0035
10,0.295,0.009,0.03,0.0349,0.0037,0.0035
20,0.293,0.009,0.029,0.0346,0.0036,0.0035
30,0.291,0.009,0.029,0.0343,0.0036,0.0034
40,0.289,0.009,0.029,0.034,0.0036,0.0034
50,0.287,0.009,0.029,0.0337,0.0035,0.0034
100,0.278,0.01,0.028,0.0323,0.0034,0.0032
200,0.26,0.013,0.026,0.0298,0.0032,0.003
300,0.245,0.014,0.024,0.0277,0.003,0.0028
400,0.231,0.015,0.023,0.0259,0.0028,0.0026
500,0.219,0.015,0.022,0.0244,0.0026,0.0024
\end{filecontents*}
\csvreader[head to column names, late after line=\\]{Li8TlabTable.csv}{}
{\ \Elab 
&
\SI{\sigmaA+-\dsigmaAA}{}
& 
\SI{\sigmaB+-\dsigmaBA}{}
}
\end{tabular}
\end{ruledtabular}
%\end{adjustbox}
 \label{table:Li8TlabTable}
\end{table}  

%%------------- Figure-6 --------------------------------
\begin{figure}[tbh]
\begin{center}
\includegraphics[width=0.47\textwidth,clip=true]{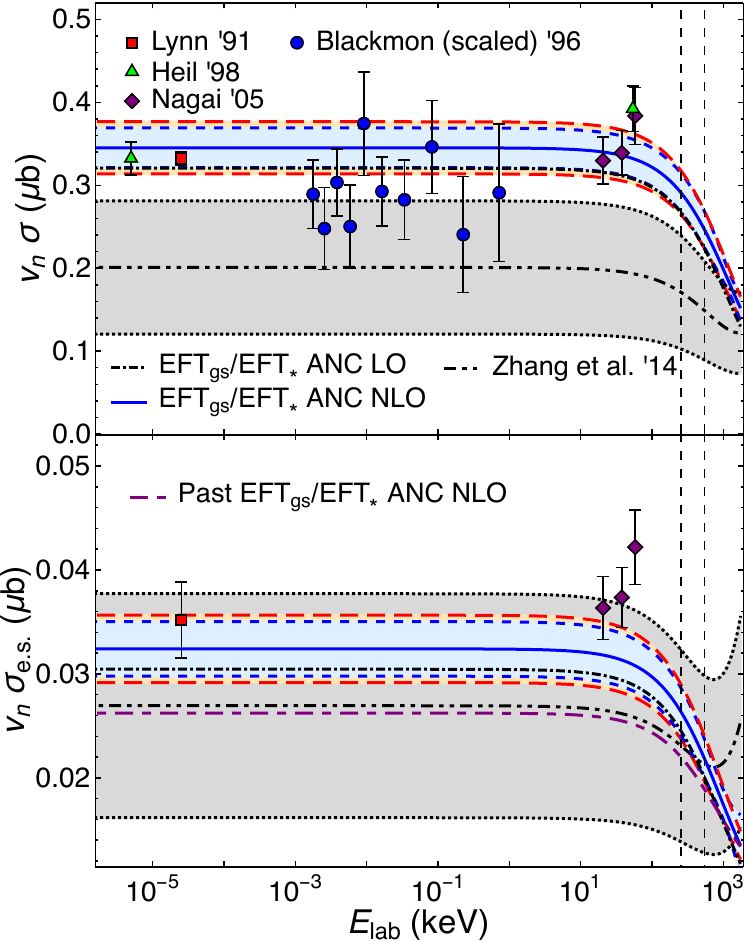}
\end{center}
\caption{\protect  Top panel: Total capture to the $2^+$ and $1^+$ state, bottom panel capture to the $1^+$ state. These results in the lab frame were constrained by the measured ANCs~\cite{Trache:2003ir}.
The theory curves, shaded regions, grid lines and data legends have the same meaning as in Fig.~\ref{fig:TotalCapture}. 
The (purple) long-short-dashed curve is the NLO capture to the $1^+$ state with the wrong Clebsch-Gordan coefficients~\cite{Fernando:2011ts}. 
 Ref.~\cite{Nagai:2005} measured both the capture and the branching ratio [\num{0.89+-0.01}] to the $2^+$ state which we used to plot the capture data for the $1^+$ state.   The (black) dot-dot-dashed curve is the result by Zhang \emph{et al.}~\cite{Zhang:2013kja}. The (gray) shaded region between the (black) dotted lines indicate the 40\% theory error estimated by Zhang \emph{et al.}  The text provides more context for comparison of the current work with Ref.~\cite{Zhang:2013kja}. }
\label{fig:TotalCaptureANC}
\end{figure}

As alternative fitting procedure one determines the unknown couplings 
from just the ANCs, like the last two rows of 
Tables~\ref{table:Li8Table} and ~\ref{table:Li8TableB}.
The thermal ratios at LO and NLO, which are predictions in this case, show a converging pattern consistent with the assumed $Q/\Lambda\sim 1/3$ estimate. 
In Fig.~\ref{fig:TotalCaptureANC} the total capture rate (top panel) and capture rate to the excited state (bottom panel) of $^8$Li using the measured ANCs~\cite{Trache:2003ir} as input are shown. The LO  (dot-dashed curve) and NLO (solid curve) result show a convergence pattern, and the NLO result  is well within the estimated 30\% correction to the LO result. We include the expected 10\% NNLO theory error to the NLO curve. The errors from the input parameters are dominated by the ANC errors, and are nearly as large as the theory errors. We also show, in the bottom panel, the capture to the excited state using the expression from earlier calculation~\cite{Fernando:2011ts} that had a mistake as the long-short-dashed curve. The corrected expression in Eq.~(\ref{eq:1plusstate}) with $|b|^2=5/6$ instead of $|b|^2=1/2$, shown as the solid curve, gives a more accurate description of thermal data~\cite{Lynn:1991}. In Ref.~\cite{Fernando:2011ts}, $\mathcal Z^{(^3P_2)}$ was fitted to thermal capture, and so the mistake would not have been apparent.

In Fig.~\ref{fig:TotalCaptureANC}, we also show the calculation by Zhang \emph{et al.}~\cite{Zhang:2013kja} by the dot-dot-dashed curve. We have indicated the 40\% theory error estimated in Ref.~\cite{Zhang:2013kja}. One should note, however, that Zhang \emph{et al.} includes at LO capture from both $S=2$ $s$-wave initial state with strong interaction and $S=1$ $s$-wave initial state without strong interaction. This makes their result, in our power counting, NLO that includes the same exact contributions to the capture cross section as ours.  Their result also includes capture from initial $d$-wave without strong interaction which we count as NNLO. Thus we would estimate their theory error to be less than 40\%.  The use of the measured ANCs~\cite{Trache:2003ir} in Fig.~\ref{fig:TotalCaptureANC} instead of the calculated ANCs used by Zhang \emph{et al.}~\cite{Zhang:2013kja}  has very little impact on the numerical value of the cross section. It seems the difference between our results (central values), particularly capture to the dominant $2^+$ ground state,  has to do with the evaluation of the wave function renormalization constants and their relations to the ANCs. Ref.~\cite{Zhang:2013kja}  expands the ANCs as LO, NLO whereas we treat them as experimental input without expansion. The LO, NLO expansion in our analysis is for theory parameters, not experimental inputs such as thermal capture rates and ANCs. Further, Zhang \emph{et al.}  do not indicate the factors $a$, $b$ that appear in Eqs.~(\ref{eq:2plusstate}),~(\ref{eq:1plusstate}) in their cross section making it difficult to infer the composition of their bound states in terms of $p_{3/2}$, $p_{1/2}$ neutron states.
The final NLO form, as a function of momentum $p$, of the cross section that we use here is the one calculated earlier in Ref.~\cite{Rupak:2011nk} (after expansion to NLO) which was also reproduced by Zhang \emph{et al.}~\cite{Zhang:2013kja}. The capture to the excited state by Zhang \emph{et al.} is found to be closer to our previous calculation~\cite{Fernando:2011ts} that has been corrected now. 
Ref.~\cite{Zhang:2013kja} calculates the $S=2$ thermal branching ratio for the $2^+$ state  as \num{0.93+-0.02} in agreement,  but as \num{0.75+-0.07} for the $1^+$ state  in disagreement with our results,  Tables~\ref{table:Li8Table},~\ref{table:Li8TableB}.

\begin{figure}[tbh]
\begin{center}
\includegraphics[width=0.49\textwidth,clip=true]{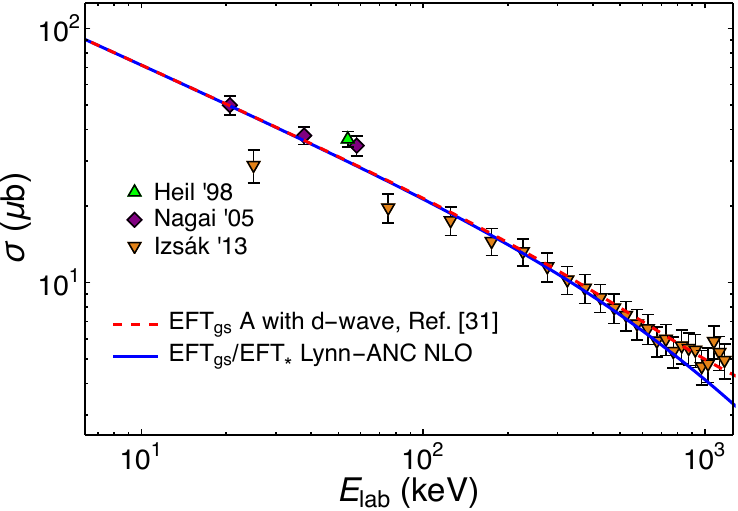}
\end{center}
\caption{\protect Capture cross section via Coulomb dissociation in the lab frame. Izs\'ak data were extracted from Ref.~\cite{Izsak:2013hga} as explained in the text. The dashed (red) curve is the \EFTg ~A result from Fig.~\ref{fig:Survey} together with capture from $d$-wave initial state~\cite{Rupak:2011nk,Fernando:2011ts}. Solid (blue) curve is the NLO \EFTg/EFT$_\star$ result from Fig.~\ref{fig:TotalCapture}. }
\label{fig:Izsak}
\end{figure}

Finally in Fig.~\ref{fig:Izsak}, we consider some recent Coulomb dissociation data that was used to predict the capture cross section. We extracted the experimental results digitally from Fig. 10 of Ref.~\cite{Izsak:2013hga}. We assumed the lowest horizontal-axis tick mark on the log-log plot to be at 
10 keV (instead of 1 keV) which gives the expected results for the known Nagai~\cite{Nagai:2005} data. We estimate our digital extraction to introduce errors of about a 1\%.  The dashed (red) curve is the \EFTg ~A result from Fig.~\ref{fig:Survey} with  the capture from initial $d$-wave states included that was published earlier~\cite{Rupak:2011nk,Fernando:2011ts}. This is not a complete NNLO  calculation in the \EFTg  ~as it lacks a two-body current contribution and $S=2$ channel $s$-wave effective range correction. However, one can see that the expected NNLO corrections to the NLO \EFTg/EFT$_\star$ solid (blue) curve from Fig.~\ref{fig:TotalCapture} moves the theory result in better agreement with the data. The difference between the dashed and solid curve is about
4\% at $E_\text{lab}\approx 1$ MeV, which is consistent with a NNLO correction. 
 
 %%------------- Figure-7 --------------------------------

 %==================================================
 \section{Conclusions}
 \label{sec:Conclusions}
 %==================================================
 
 We considered the E1 contribution to the \nLi~capture reaction at low energies. A coupled-channel calculation for the contribution from the excited $^7\mathrm{Li}^\star$ core was presented. This theory was compared to an earlier calculation which did not include the excited core as an explicit degree of freedom~\cite{Rupak:2011nk,Fernando:2011ts}. We developed a power counting for both the EFTs---one with explicit excited $^7\mathrm{Li}^\star$ core EFT$_\star$ and one without \EFTg---and show that the two theories are equivalent in their momentum-dependence up to NLO. This confirms the expectation that the $^7\mathrm{Li}^\star$ contribution to the momentum dependence,  which participates only in the sub-dominant spin $S=1$ channel with 
a branching ratio of $\sim 0.2$,  
is a higher-order effect~\cite{Rupak:2011nk,Fernando:2011ts}. Though the momentum dependence in both EFTs at this order is the same, the interpretation of the cross section in the $S=1$ channel in terms of $p$-wave effective momenta in the two theories is different. In \EFTg, we can relate the cross section to an overall normalization in terms of the $^3P_2$ effective momentum. No such simple interpretation is possible in EFT$_\star$. 

The survey presented in Sec.~\ref{sec:Survey} relaxed the 
simplification 
${r_1^{(^3P_2)}=r_1^{(^5P_2)}\sim\Lambda}$, for the $2^+$ $^8$Li ground state, 
done in previous 
works~\cite{Rupak:2011nk,Fernando:2011ts}. 
In \EFTg, Fig.~\ref{fig:Contours} 
shows how $r_1^{(^3P_2)}$ and $r_1^{(^5P_2)}$ are correlated to 
reproduce a 
given value for the thermal capture to the ground $2^+$ state of 
${}^8{\rm Li}$ and the branching ratio of the spin channel $S=2$. 
Alternatively, $r_1^{(^3P_2)}$ and $r_1^{(^5P_2)}$ can be fixed via the 
corresponding ANCs $C_{1,^3P_2}^2$ and $C_{1,^5P_2}^2$ calculated in 
Ref.~\cite{Nollett:2011qf} or taken from experiment~\cite{Trache:2003ir}, 
with similar numerical results. 
The differences from the simplified assumption $r_1^{(^3P_2)}=r_1^{(^5P_2)}$ are compatible with 
a NLO correction. 
In EFT$_\star$ the wave function 
renormalization constants ${\cal Z}^{(^5P_2)}$, ${\cal Z}^{(^3P_2)}$, 
and ${\cal Z}^{(^3P_2^\star)}$ can be fitted to the corresponding ANCs, 
though they are not enough to pin down the extra couplings that 
appear---while $r_1^{(^5P_2)}$ is uniquely determined from ${\cal Z}^{(^5P_2)}$, 
this is not true for the three $S=1$ effective momenta $r_{ij}$ in 
terms of ${\cal Z}^{(^3P_2)}$ and ${\cal Z}^{(^3P_2^\star)}$. Similar analysis for the $p$-wave channels corresponding to the $1^+$ $^8$Li excited state holds. 

We present a NLO calculation in both the EFTs with and without explicit excited $^7\mathrm{Li}^\star$ contributions. Once the wave function renormalization constants are fitted to capture data or ANCs or some combination of these, the \EFTg ~and EFT$_\star$ results are numerically the same at this order. The kinematical impact of the $^7\mathrm{Li}^\star$ core becomes important only at NNLO.
We estimate the theoretical errors in our calculation to be about 
10\% from $Q^2/\Lambda^2$ NNLO corrections. 

At NNLO, there are several contributions to the capture to the $2^+$ ground state of $^8$Li. 
Two-body current contribution to capture from the $^5S_2$ channel scales as $k_0 a_0^{(2)} L_2$~\cite{Higa:2016igc}. 
This is a $Q^2/\Lambda^2$ NNLO contribution for a natural-sized 
coupling $L_2\sim 1$, given the scalings for the scattering 
length $a_0^{(2)}\sim1/Q$ and the photon energy 
$k_0=(p^2+\gamma^2_0)/(2\mu)\sim Q^3/\Lambda^2$. 
The NNLO contributions from the $S=1$ channel start with initial state 
$s$-wave strong interactions. In \EFTg, the $s$-wave interactions can be parameterized by the scattering 
length $a_0^{(1)}$. In EFT$_\star$, 
the initial state $s$-wave interactions involve a coupled-channel 
calculation parameterized by the three scattering lengths 
$a_{11}$, $a_{12}$, $a_{22}$. Capture cross section from $d$-wave that scales as $p^4/(p^2+\gamma_0^2)^2$ was included in Refs.~\cite{Rupak:2011nk,Fernando:2011ts}. However, it is kinematically suppressed at low momentum, contributing around 10\% at $p\gtrsim 40$ MeV. So this contribution can also be included as NNLO in the dominant spin  $S=2$ channel. 
In the same spin channel the initial state ${}^5S_{2}$ receives a NNLO correction proportional to $[a_0^{(2)}r_0^{(2)}p^2/2]^2$  while the shape parameter ${\cal P}_0^{(2)}$ enters at N${}^3$LO~\cite{Chen:1999tn}. 
The NNLO contributions for the capture to the $1^+$ excited state of $^8$Li are similar to those discussed for the $2^+$ ground state. 

The EFT formalism and the theory expressions for the cross section in this work,
with the excited $^7\mathrm{Li}^\star$ core contributions, are different from those in Ref.~\cite{Zhang:2013kja}. We also differ in the interpretation of the wave function renormalization constants in terms of the $p$-wave scattering parameters. Given that the excited $^7\mathrm{Li}^\star$ core contributions to the momentum dependence of the cross section are a NNLO effect, one would expect the numerical results of Ref.~\cite{Zhang:2013kja} to be similar to those obtained here, which in turn do not differ significantly (differences $\lesssim 1.5\%$ for $p\lesssim 40$ MeV) from earlier calculations (excluding $d$-wave contributions) in Refs.~\cite{Rupak:2011nk,Fernando:2011ts}. 
However, Fig.~\ref{fig:TotalCaptureANC} shows a considerable numerical discrepancy between our and Zhang \emph{et al.}'s~\cite{Zhang:2013kja} results. Since Fig.~\ref{fig:TotalCaptureANC} uses the same ANCs as input,  the discrepancy  has to do with the relation between the EFT couplings and the ANCs. 

In this work, we included only the non-resonant capture. Future work would include the M1 contribution from the $3^+$ 
resonance to \nLi~\cite{Higa:2020}. The NNLO E1 and LO M1 contributions to \pBe ~has recently been calculated using the current coupled-channel  formalism. The  excited $^7\mathrm{Be}^\star$ contribution has a kinematical impact for energy ${E\gtrsim 500}$ keV at NNLO~\cite{Higa:2020pvj}.

%=============================x======================
\begin{acknowledgments}
We thank C. Bertulani and A. Horv\'ath for discussing their work on Coulomb dissociation with us. We benefited from discussions with K. M. Nollett, D. R. Phillips, and X. Zhang.
This work was supported in part by 
U.S. NSF grants PHY-1615092 and PHY-1913620 (PP, GR) 
and Brazilian agency FAPESP thematic projects 2017/05660-0 and 2019/07767-1, and INCT-FNA Proc. No. 464898/2014-5 (RH).
The cross section figures for this article have been created using SciDraw~\cite{SciDraw}. 
\end{acknowledgments}

\appendix
%==========================================
\section{Projectors}
\label{sec:Projectors}
%===========================================

The following are from Ref.~\cite{Fernando:2011ts} that we include for reference. 
For each partial wave we construct the corresponding projection operators 
from the relative core-nucleon velocity, the spin-1/2 
Pauli matrices $\sigma_{i}$'s, and the following spin-1/2 to spin-3/2 
transition matrices 
\begin{align}
S_1&=\frac{1}{\sqrt{6}}\left(\begin{array}{cccc}
-\sqrt{3} & 0 & 1 & 0\\
0&-1&0&\sqrt{3}
\end{array}\right)\,, 
\nonumber\\
S_2 &= -\frac{i}{\sqrt{6}} \left(\begin{array}{cccc}
\sqrt{3} & 0 & 1 & 0\\
0&1&0&\sqrt{3}
\end{array}\right)\,, 
\nonumber\\
S_3 &= \frac{2}{\sqrt{6}} \left(\begin{array}{cccc}
0 & 1 & 0 & 0\\
0&0&1&0
\end{array}\right)\,,
\end{align}
which satisfy
\begin{align}
S_{i}S^{\dagger}_{j}&=\frac{2}{3}\delta_{ij}-\frac{i}{3}\epsilon_{ijk}
\sigma_{k}\,,
\nonumber\\
S_{i}^{\dagger}S_{j}&=\frac{3}{4}\delta_{ij}-\frac{1}{6}\big\{J_{i}^{(3/2)},
J_{j}^{(3/2)}\big\}+\frac{i}{3}\epsilon_{ijk}J_{k}^{(3/2)}\,,
\end{align}
where $J_{i}^{(3/2)}$'s are the generators of the spin-3/2. 
We construct the Clebsch-Gordan coefficient matrices 
\begin{align}
F_i &=-\frac{i\sqrt{3}}{2}\sigma_2 S_i\, , &
Q_{i j} &= -\frac{i}{\sqrt{8}}\sigma_2\big(\sigma_i S_i+\sigma_j S_i\big),
\label{eq:ncorespinmatrices}
\end{align}
for projections onto spin channels $S=1$ and $S=2$, respectively. Then in coordinate space the relevant projectors that appear in the Lagrangians involving the $^7$Li ground state in Eqs. (\ref{eq:EFT}), (\ref{eq:EFTstar}) are~\cite{Rupak:2011nk,Fernando:2011ts}
\begin{align}
P_i^{(^3S_1)} &= F_j\, , 
\nonumber\\
P_{ij}^{(^5S_2)} &= Q_{ij}\,,
\nonumber\\
P^{(^3P_1)}_i &=\sqrt{\frac{3}{2} }F_x\P_y \epsilon_{i x y}\,, 
\nonumber\\
P^{(^3P_2)}_{i j} &=\sqrt{3} F_x \P_y\ R_{x y i j }\,,
\nonumber\\
P^{(^5P_1)}_i &=\sqrt{\frac{9}{5}} Q_{i x}\P_ x\,, 
\nonumber\\
P^{(^5P_2)}_{i j}&=\frac{1}{\sqrt{2}} Q_{x y} \P_z\ T_{x y z i j}\, . 
\label{eq:projdefr-space}
\end{align}
The tensors 
\begin{align}
R_{ijxy}&=\frac{1}{2}\left(\delta_{i x}\delta_{j y}+\delta_{i y}\delta_{j x} 
-\frac{2}{3}\delta_{i j}\delta_{x y}\right), 
\nonumber\\
T_{xyz i j}&=\frac{1}{2}\Big(\epsilon_{x z i}\delta_{y j}
+\epsilon_{x z j}\delta_{y i}
+\epsilon_{y z i}\delta_{x j}+\epsilon_{y z j}\delta_{x i} \Big),
\label{eq:RTGtensorsdef}
\end{align}
ensures total angular momentum $J=2$ is picked. 

The new projectors to describe the interactions in Eq.~(\ref{eq:EFTstar}) with the excited $^7\mathrm{Li}^\star$ core are 
\begin{align}\label{eq:ProjectorsEFTstar}
     P_i^{(^3S_1^\star)} & = -\frac{i}{\sqrt{2}}\sigma_2\sigma_i \, , & &\nonumber\\
    P_i^{(^3P_1^\star)} & =  -i\frac{\sqrt{3}}{2}\sigma_2 \sigma_x \P_y \epsilon_{i x y}\, , 
    \nonumber\\
    P_{ij}^{(^3P_2^\star)} &= -i\sqrt{\frac{3}{2}}\sigma_2\sigma_x \P_y R_{xyij}\, .
\end{align}

For the external states we introduce the photon vector ($\varepsilon^{(\gamma)}_{i}$), excited state $^8$Li $1^+$ 
spin-1 ($\varepsilon_{j}$), and  ground state $^8$Li $2^+$ spin-2 ($\varepsilon_{ij}$) polarizations, obeying the 
following polarization  sums~\cite{Choi:1992,Fleming:1999ee}, 
\begin{align}
\sum_{\rm pol.}\varepsilon^{(\gamma)}_{i}\varepsilon^{(\gamma)*}_{j}&=
\delta_{ij}-\frac{k_ik_j}{k^2}\,, 
\nonumber\\
\sum_{\rm pol.\ ave.}\varepsilon_{i}\varepsilon^{*}_{j}
&=\frac{\delta_{ij}}{3}\,, 
\nonumber\\
\sum_{\rm pol.\ ave.}\varepsilon_{ij}\varepsilon^{*}_{lm}
&=\frac{R_{ijlm}}{5}\,.
\label{eq:poltensordefs}
\end{align}

%% Breaks long references 
\setcounter{biburlnumpenalty}{9000}
\setcounter{biburllcpenalty}{7000}
\setcounter{biburlucpenalty}{8000}

\bibliographystyle{apsrev4-2}
%\bibliography{Reference.bib}

\end{document}